\documentclass[%
 reprint,
 amsmath,amssymb,
 aps,
superscriptaddress,
]{revtex4-2}

\usepackage[dvipsnames]{xcolor}
\usepackage{graphicx}
\usepackage{dcolumn}
\usepackage{bm}
\usepackage{lipsum}
\usepackage{amsmath}

\usepackage{url}
\usepackage[breaklinks]{hyperref}

\usepackage[dvipsnames]{xcolor}
\definecolor{mblue}{RGB}{42, 54, 144} 
\hypersetup{colorlinks=true,breaklinks=true,urlcolor=blue,linkcolor=blue,citecolor=blue}

\usepackage{placeins}
\usepackage[mathlines]{lineno}
\usepackage{siunitx}
\usepackage{braket}

\DeclareSIUnit\gauss{G}
\usepackage{tikz}

\definecolor{verdealberto}{rgb}{0.05, 0.64, 0.09}

\definecolor{chunkit_color}{rgb}{0.55, 0.35, 0.46}

\begin{document}

\title{Optically trapped Feshbach molecules of fermionic $^{161}$Dy and $^{40}$K:\\ Role of light-induced and collisional losses}

\author{Alberto Canali}
\affiliation{Institut f{\"u}r Experimentalphysik, Universit{\"a}t Innsbruck, Austria}

\author{Chun-Kit Wong}
\affiliation{Institut f{\"u}r Experimentalphysik, Universit{\"a}t Innsbruck, Austria}
\affiliation{Institut f{\"u}r Quantenoptik und Quanteninformation (IQOQI), {\"O}sterreichische Akademie der Wissenschaften, Innsbruck, Austria}

\author{Luc Absil}
\affiliation{Institut f{\"u}r Experimentalphysik, Universit{\"a}t Innsbruck, Austria}

\author{Zhu-Xiong~Ye}
\altaffiliation[Current address: ]{State Key Laboratory of Quantum Optics Technologies and Devices, Institute of Opto-Electronics, Shanxi University, Taiyuan, China.}
\affiliation{Institut f{\"u}r Experimentalphysik, Universit{\"a}t Innsbruck, Austria}

\author{Marian Kreyer}
\affiliation{Institut f{\"u}r Experimentalphysik, Universit{\"a}t Innsbruck, Austria} 
\affiliation{Institut f{\"u}r Quantenoptik und Quanteninformation (IQOQI), {\"O}sterreichische Akademie der Wissenschaften, Innsbruck, Austria}

\author{Emil Kirilov}
\affiliation{Institut f{\"u}r Experimentalphysik, Universit{\"a}t Innsbruck, Austria} 
\affiliation{Institut f{\"u}r Quantenoptik und Quanteninformation (IQOQI), {\"O}sterreichische Akademie der Wissenschaften, Innsbruck, Austria}

\author{Rudolf Grimm}
\affiliation{Institut f{\"u}r Experimentalphysik, Universit{\"a}t Innsbruck, Austria} 
\affiliation{Institut f{\"u}r Quantenoptik und Quanteninformation (IQOQI), {\"O}sterreichische Akademie der Wissenschaften, Innsbruck, Austria}

\date{\today}

\begin{abstract}
We study the decay of a dense, ultracold sample of weakly bound DyK dimers stored in an optical dipole trap.
Our bosonic dimers are composed of the fermionic isotopes $^{161}$Dy and $^{40}$K, which is of particular interest for experiments related to pairing and superfluidity in fermionic systems with mass imbalance.
We have realized dipole traps with near-infrared laser light in four different wavelength regions between 1050 and 2002 \unit{nm}. We have identified trap-light-induced processes as the overall dominant source of losses, except for wavelengths around 2000 \unit{nm}, where light-induced losses appeared to be much weaker. In a trap near 1550 \unit{nm}, we found a plateau of minimal light-induced losses, and by carefully tuning the wavelength we reached conditions where losses from inelastic collisions between the trapped dimers became observable. For very weakly bound dimers close to the center of a magnetically tuned Feshbach resonance, we demonstrate the Pauli suppression of collisional losses by about an order of magnitude.

\end{abstract}

\maketitle

\section{Introduction}

Since its first demonstration almost 40 years ago \cite{Chu1986eoo}, the trapping of atoms by the optical dipole force in off-resonant laser light \cite{Grimm2000odt} has become a key technique in modern atomic physics and quantum sciences \cite{Wolswijk2025tma}. Optical dipole traps (ODTs) have found numerous applications, e.g.\ for ultraprecise atomic clocks \cite{Ludlow2015oac, Yang2025cbb}, for precision measurements \cite{Zheng2022mot, Thekkeppatt2025mot, Ravensbergen2018ado}, and for quantum computation and simulation \cite{Browaeys2020mbp,Kaufman2021qsw, Anand2024ads, Gruen2024ota}. In the broad research field of ultracold quantum gases \cite{Varenna1998book, Varenna2006book, Varenna2022book}, optical dipole trapping schemes have, in many cases, served as the enabling tool to achieve quantum degeneracy \cite{Weber2003bec, Takasu2003ssb, Griesmaier2005bec} and to create novel quantum-degenerate systems with single species, spin mixtures, and mixtures of different species.

In the rapidly developing field of ultracold molecules, ODTs are widely employed for the preparation of cold and dense atomic samples, serving as a starting point to create molecules by subsequent photo- \cite{Jones2006ups, Deiglmayr2008fou} or magneto-association \cite{Koehler2006poc, Chin2010fri} techniques. A multitude of different experiments have been carried out on optically trapped molecules, such as collisional studies \cite{Chin2005oof, Ferlaino2008cbt, Ni2010dco} and the manipulation of internal states \cite{Mark2007sou, Winkler2007cot, Lang2008ctm}. A very prominent example is the transfer of dimers to their rovibronic ground state \cite{Ni2008ahp, Lang2008utm, Danzl2010auh}.
Quantum degeneracy of molecular samples was achieved more than 20 years ago with the observation of molecular Bose-Einstein condensation (BEC) of weakly bound pairs of fermionic atoms \cite{Jochim2003bec, Greiner2003eoa, Zwierlein2003oob} and the experimental realization of the crossover from BEC to Bardeen-Cooper-Schrieffer (BCS) type systems \cite{Bartenstein2004cfa, Regal2004oor, Zwierlein2004cop, Zwerger2012tbb, Strinati2018tbb}. The condensation of weakly bound molecules made of bosonic atoms was demonstrated in Ref.~\cite{Zhang2021tfa} and, most recently, BEC of ground-state dipolar molecules was achieved \cite{Bigagli2024oob,Shi2025bec}. Another important development in the field is the optical dipole trapping and manipulation of diatomic \cite{Zhang2020fas, Cairncross2021aoa, Ruttley2023fou} and polyatomic \cite{Anderegg2019aot,Vilas2024aot} molecules using optical tweezers and tweezer arrays. 
To fully exploit the wide application potential of ODTs for experiments on ultracold molecular samples, thorough understanding of the limiting processes is required. The complex interactions between molecules along with their very rich level structure can lead to additional effects not relevant for trapped atomic samples.

In our laboratory, we have demonstrated \cite{Soave2023otf} the preparation of optically trapped samples of weakly bound bosonic dimers, composed of two different fermionic constituents, $^{161}$Dy and $^{40}$K. We have achieved efficient molecule production by applying standard magneto-association techniques (`Feshbach ramps' \cite{Koehler2006poc, Chin2010fri}) to a double-degenerate atomic mixture. The samples of `Feshbach molecules' created in this way are of great interest in view of molecular Bose-Einstein condensation in heteronuclear systems and, in a more general sense, as a starting point for the realization of many-body states of strongly interacting fermions with mass imbalance in BEC-BCS crossover regimes \cite{Ciamei2022ddf, Ravensbergen2020rif, Ye2025dms}. Such systems are particularly promising for the realization of asymmetric superfluids with unconventional pairing mechanisms~\cite{Gubbels2009lpi, Gubbels2013ifg, Wang2017eeo, Pini2021bmf}, most notably the elusive Fulde-Ferrell-Larkin-Ovchinnikov (FFLO) state~\cite{Fulde1964sia, Larkin1964nss, Radzihovsky2010ifr}. In addition to that, a variety of interesting few-body phenomena has been predicted to emerge in resonant fermion mixtures \cite{Naidon2017epa, Kartavtsev2007let, Zaccanti2022mif}.

A question of central importance concerns the stability of the trapped dimer sample against inelastic dimer-dimer collisions. For homonuclear dimers in fermionic spin mixtures, it is well known that a Pauli suppression effect \cite{Petrov2004wbd, Varenna2006book} can strongly reduce collisional losses, thus greatly enhancing the stability of the system. We would expect a similar effect to be present for our weakly bound DyK molecules \cite{Petrov2005dmi, Jag2016lof}, but our previous attempts to study collisional effects \cite{Soave2023otf} were obstructed by another, by far dominant loss mechanism. Our experiments provided strong evidence for a light-induced decay caused by the ODT itself. The loss effect has been noted before in a few bi-alkali molecular systems~\cite{Chotia2012lld, Zhang2020fas, Spence2023ASR}, but it appears to play a much more important role for more complex dimers like LiCr~\cite{Finelli2024ula} and DyK.

In the present work, we report on systematic studies of losses from an optically trapped, dense sample of weakly bound DyK molecules. After summarizing the main properties of the dimers in the vicinity of a Feshbach resonance (Sec.~\ref{Sec:FeshbachMol}) and outlining the main experimental procedures and conditions in (Sec.~\ref{Sec:SamplePrep}), we present detailed measurements of trap-light-induced losses (Sec.~\ref{Sec:LighInducedLosses}). We investigate four different wavelength regions in the near-infrared, carrying out spectroscopic wavelength scans in certain regions of interest and measuring the linear coefficient that characterizes intensity-dependent losses. We also identify regions of minimal losses. With a proper choice of the trap wavelength mitigating the detrimental effect of light-induced losses, we identify the losses caused by inelastic dimer-dimer collisions (Sec.~\ref{Sec:Collisions}) and we measure the corresponding rate coefficient as a function of the magnetic detuning from resonance. We indeed observe a reduction of collisional losses close to resonance by about an order of magnitude, which is in accordance with theoretical expectations for the Pauli suppression effect. Finally (Sec.~\ref{Sec:Conclusions}), we discuss the implications of our findings for future experiments.

\section{D\lowercase{y}K Feshbach molecules}
\label{Sec:FeshbachMol}

As our essential tool to control the $s$-wave interaction between $^{161}$Dy and $^{40}$K atoms and to form molecules, we employ a low-field interspecies Feshbach resonance near \SI{7.3}{G} \cite{Ye2022ool, Soave2023otf}. The narrow resonance is well isolated from other interspecies resonances and can thus be described in terms of a standard two-channel model well established in the literature \cite{Petrov2004tbp, Chin2010fri}. Moreover, the resonance is not contaminated by intraspecies resonances \cite{Ye2022ool}. To set the stage, we summarize the main properties of this resonance and the relevant parameter values, which were accurately determined in our previous work \cite{Ye2022ool, Soave2023otf}.

\begin{figure}[]
    \centering
    \includegraphics[width=1\columnwidth]{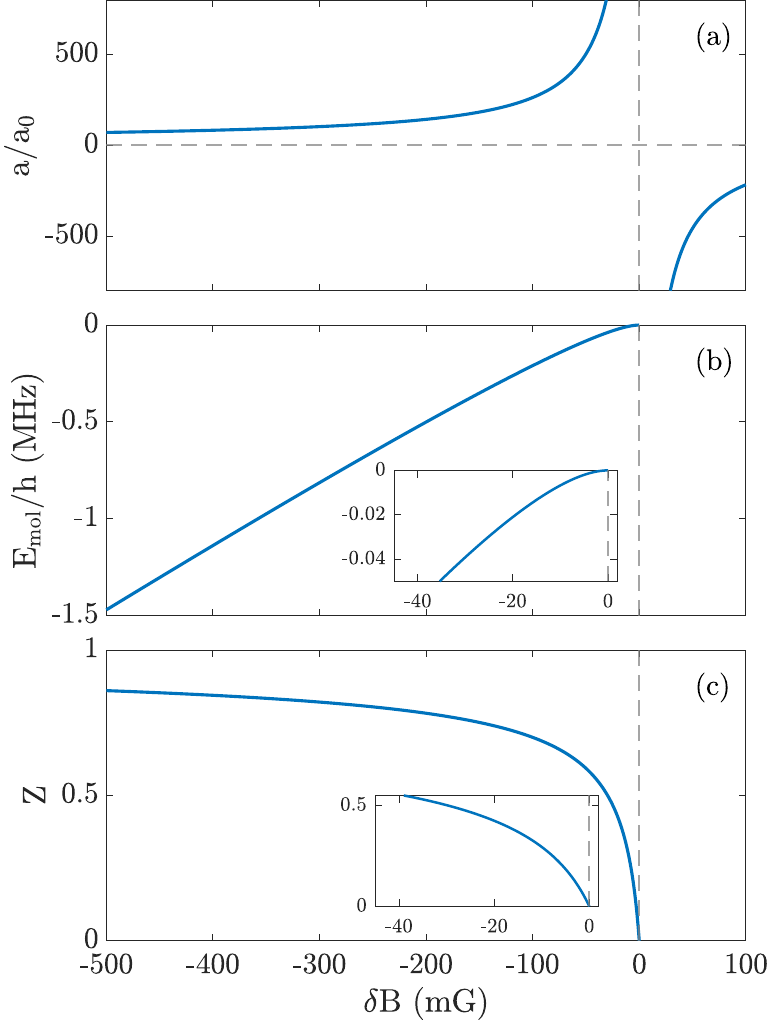}
    \caption{Properties of the 7.3-G Feshbach resonance. (a)~Scattering length $a$, (b) molecular state energy $E_{\rm mol}$, and (c) closed-channel fraction $Z$ as functions of the magnetic detuning $\delta B$. The insets in (b) and (c) focus on the near-universal region close to resonance.}
    \label{fig:FeshbachRes}
\end{figure}

Close to the resonance, the $s$-wave scattering length can be expressed as 
\begin{equation}\label{Eq:scatteringLength}
a(B)= a_{\rm bg}- \frac{A}{\delta B}\, a_0 ,
\end{equation}
where $A$ characterizes the strength of the resonance, $\delta B=B-B_0$ represents the magnetic detuning from the resonance center, $a_{\rm bg}$ denotes the background scattering length, and $a_0$ is the Bohr radius. To our best knowledge, resonance parameters are $B_0=\SI{7.276(2)}{G}$, $A=\SI{24.0(6)}{G}$, and $a_{\rm bg}=23(5)\,a_0$ \cite{Ye2022ool,Soave2023otf}. In Fig.~\ref{fig:FeshbachRes}(a), we show the scattering length in proximity to the resonance.

In the case of a `narrow' Feshbach resonance, also referred to as `closed-channel dominated resonance' \cite{Chin2010fri}, it is necessary to introduce an additional parameter to fully characterize the interaction properties. For this purpose, it is convenient to consider the range parameter $R^*=\hbar^2/(2m_ra_0\delta\mu A)$ \cite{Petrov2004tbp}, where $m_{\rm r}$ is the reduced mass and $\delta\mu=\mu_{\rm open}-\mu_{\rm closed}$ is the difference in magnetic moments between the atomic scattering state (open channel) and the molecular state underlying the Feshbach resonance (closed channel). For $a\gtrsim R^*$, the interaction physics  acquires universal behavior in the sense that a single parameter, such as the scattering length, is sufficient to describe the underlying physics \cite{Braaten2006uif}. For the resonance considered here $R^*=604(20)\, a_0$, which corresponds to a universal regime accessible for $|\delta B|\lesssim\SI{40}{mG}$.

The molecular state energy related to a narrow Feshbach resonance can be expressed as
\begin{equation}\label{Eq:BindingEnergy}
    E_{\mathrm{mol}}(B)=-\delta\mu\,\delta B^*\left(\sqrt{1-\frac{\delta B}{\delta B^*}}-1 \right)^2.
\end{equation}
Here, for brevity, we have introduced the magnetic field scale $\delta B^*=m_r\,\delta\mu\,a_0^2\,A^2/(2\hbar^2)$, for which $a(-\delta B^*)=4R^*$, which corresponds to $\delta B^*=\SI{9.9(6)}{mG}$, in our case. In Fig.~\ref{fig:FeshbachRes}(b), we show the binding energy of the DyK dimers according to Eq.~\eqref{Eq:BindingEnergy}, in the magnetic detuning range 
$-\SI{500}{mG} < \delta B < 0$.

The closed-channel fraction quantifies the admixture between the bare molecular state and the free-atom scattering state in the wavefunction of the Feshbach molecules. It can be obtained from the molecular binding energy through the differential magnetic moment, $\delta \mu_{\rm mol}(B)=\partial E_{\mathrm{mol}}/\partial B$. The closed-channel fraction is then given by 
\begin{equation}\label{Eq:ClosedChannelFraction}
Z(B)=\delta\mu_{\mathrm{mol}}(B)/ \delta\mu,
\end{equation}
 and Fig.~\ref{fig:FeshbachRes}(c) shows the resulting $Z(B)$ for DyK molecules near the 7.3-G resonance.

\section{Experimental Sequence}
\label{Sec:SamplePrep}
 
The preparation of the molecular sample closely follows the procedures developed in our previous work \cite{Ravensbergen2018poa, Ye2022ool, Soave2023otf}. As a starting point, a double-degenerate mixture of $^{161}$Dy and $^{40}$K is prepared with both species spin-polarized in their lowest hyperfine states, $\ket{F,m_F}=\ket{21/2,-21/2}$ and $\ket{9/2,-9/2}$, respectively. 
{Dimers formed in the lowest hyperfine spin channel cannot dissociate spontaneously because of the absence of energetically lower atomic pair states. Therefore, 
our system is immune against potentially fast dissociation losses as studied for homo- and heteronuclear Feshbach molecules in Refs.~\cite{Thompson2005sdo, Jag2016lof}.
}

We first prepare the double-degenerate atomic mixture at low magnetic field and then ramp the field to the vicinity of the \SI{7.3}{G}-Feshbach resonance. DyK molecules are subsequently associated by performing a magnetic field sweep across this resonance \cite{Chin2010fri, Soave2023otf}. The molecular sample obtained by this Feshbach ramp is then purified by removing the remaining atoms using a Stern-Gerlach technique \cite{Soave2023otf}. 
With this procedure, we produce a pure molecular sample of typically \SI{10000}{} DyK molecules at a temperature of around \SI{70}{nK}, with mean trap frequency  $\bar\omega=2\pi\times\SI{31}{Hz}$. Compared with our previous work, we are now able to obtain twice the number of molecules. We achieve this improvement by carefully optimizing both the mixture preparation sequence and  the Stern-Gerlach purification scheme. In particular, we perform the cleaning sequence at a detuning of $\delta B=\SI{-40}{mG}$, which is \SI{80}{mG} closer to resonance than previously, and reduces light-induced losses of molecules during the process. 

All results presented in this work are based on measuring the number of the molecules after a variable hold time in a target ODT. After the purification process, the molecular sample is transferred into the ODT under investigation, where the measurement takes place. At the end of the investigation time, the ODT is switched off and the magnetic field is quickly ramped across the resonance to positive detunings, thus dissociating the molecules. The number of molecules is finally obtained from absorption images of the resulting K atoms \footnote{The potassium imaging provides a better quality signal compared to the Dy ones, due to the  higher absorption cross section.}.

\section{Light-Induced Losses}
\label{Sec:LighInducedLosses}
We investigate the dependence of light-induced losses on wavelength  and intensity of the near-infrared light used for the ODT. In Sec.~\ref{subSec:Spectroscopy}, we report the results of spectroscopic measurements performed on the DyK Feshbach dimers in two different wavelength regions. In Sec.~\ref{subSec:LightIntensity}, we compare light-induced losses at different wavelengths by observing the dependence of loss rate on intensity.

\subsection{Spectroscopy in two regions of interest}
\label{subSec:Spectroscopy}

We perform spectroscopy on the DyK Feshbach molecules by scanning the wavelength of the near-infrared trapping light and observing losses from the ODT. For generating the trap light, we have the choice between two fiber lasers, with wavelengths centered around \SI{1051}{nm} (NKT Koheras BOOSTIK Y10, linewidth $<\,$\SI{20}{kHz}) and \SI{1547}{nm} (NKT Koheras BOOSTIK E15, linewidth $<\,$\SI{0.1}{kHz}). Both lasers feature a tuning range of around \SI{1}{nm}. 
 
\begin{figure}[tb]
\centering
\includegraphics[trim=10 7 37 18,clip,width=1\columnwidth]{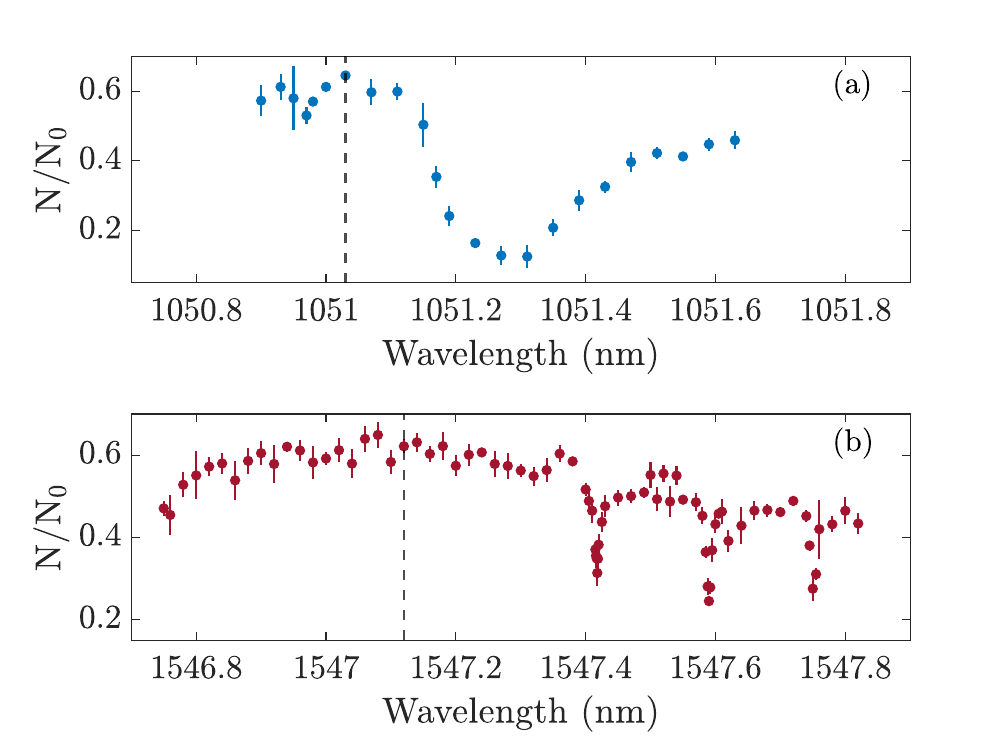}
\caption{Trap-loss spectroscopy on DyK Feshbach molecules by variation of the ODT wavelength. Panel (a) refers to the region around \SI{1051}{nm}, with initial number $N_0=\SI{1.1e4}{}$ and  a hold time of \SI{10}{ms}. Panel (b) refers to the region around \SI{1547}{nm}, with  $N_0=\SI{9e3}{}$ and a hold time of \SI{33}{ms}. The error bars represent $1\sigma$ standard errors, calculated from multiple repetitions. The vertical dashed lines in (a) and (b) indicate the selected wavelengths \SI{1051.03}{nm} and \SI{1547.12}{nm}, respectively, used for all further measurements presented in this work.}
\label{fig:Spectroscopy}
\end{figure}

The molecules are loaded within \SI{1}{ms} from a $1064$-nm crossed-beam ODT into the single-beam ODT used for spectroscopy. The molecules are then held at a  magnetic detuning of $\delta B=\SI{-60}{mG}$  for a fixed amount of time,  while being levitated by a magnetic field gradient. We choose a hold time of \SI{10}{ms} for the $1051$-nm case and \SI{33}{ms} for the $1547$-nm case, in order to obtain a similar contrast for loss features in the two signals. We report the main results of these measurements in Fig.~\ref{fig:Spectroscopy}. For the $1051$-nm case, the ODT has a waist of \SI{72}{\micro\meter} and a power of \SI{46}{mW}, corresponding to a peak intensity of \SI{0.55}{kW/cm^2}. For the $1547$-nm case, the ODT has a waist of \SI{90}{\micro\meter} and the measurements were taken using a power of \SI{220}{mW}, corresponding to a peak intensity of \SI{1.7}{kW/cm^2}.

In Fig.~\ref{fig:Spectroscopy}(a), we report the result of the trap-loss spectroscopy for the $1051$-nm case. Our main observation is a broad loss feature, showing a large width of  about $\SI{0.2}{nm}$ ($\sim\SI{60}{GHz}$) together with very fast losses. This suggests a strong, resonant coupling to the manifold of electronically excited molecular states. 
{Without knowledge of the DyK molecular structure, we can only speculate about the particular mechanism and the physical origin of the very broad loss feature. Light-induced dissociation of the molecules \cite{Jones2006ups} could result from coupling to the repulsive branch of an excited-state molecular potential. Another possibility would be a negative dynamic polarizability of the closed-channel dimer on the blue side of a molecular resonance as observed in Ref.~\cite{Melbourne}. This would lead to an anti-trapping potential for the molecules and thus to rapid trap losses.
As a feature of practical interest, we also identify a plateau with a local maximum of the molecular lifetime on the shorter wavelength side of the resonance.} 

In Fig.~\ref{fig:Spectroscopy}(b), we show a corresponding spectroscopic measurement in the region around \SI{1547}{nm}. In contrast to the 1051-nm case, we do not observe any broad resonance features. Instead, we identify three much sharper loss features on a slowly varying background. By performing individual Lorentzian fits on each of the features, we extract a typical linewidth (full width at half maximum) of approximately $\SI{1.2}{GHz}$ and a separation between neighboring lines of about \SI{20}{GHz}. This separation between the lines is of the order of typical rotational splittings in diatomic molecules, as observed in other heteronuclear systems, like KRb \cite{Kasahara1999dfo}, and NaRb \cite{Guo2016coa,Guo2017hrm}, {but the lines seem to be unusually broad. Detailed knowledge of the DyK molecular structure will be needed for a comprehensive interpretation of all these observations.}

\subsection{Dependence on light intensity}
\label{subSec:LightIntensity}

Here we investigate the dependence of the observed light-induced losses on the intensity of the trapping light. For the two wavelength regions discussed before, we avoid loss resonances by choosing the specific wavelengths where we have identified plateaus with minimal losses, at 1051.03 nm and 1547.12 nm (vertical dashed lines in Fig.~\ref{fig:Spectroscopy}). In addition to these two wavelengths, we operate an ODT based on a single-frequency fiber laser at \SI{2002}{nm} (Precilaser FL-SF-2001-20-CW, linewidth $\sim~\SI{5}{kHz}$), and we reanalyze our earlier measurements with a single-frequency laser at \SI{1064.04}{nm} (Azurlight ALS-IR-1064-50-I-CC-SF, linewidth $<\,\SI{50}{kHz}$) \cite{Soave2023otf}. The experiments with four different laser sources allow us to investigate and compare the loss behavior in four different near-infrared wavelength regions.

The 1064-nm laser source is not tunable and therefore does not allow for a search of points of minimal losses. This makes the choice of the wavelength a matter of chance, but with a reasonable probability to avoid detrimental resonances. Regarding the 2002-nm light, we performed similar wavelength scans as the ones near \SI{1051}{nm} and \SI{1547}{nm}, but interference effects in our optical set-up, which was not designed for this wavelength, masked the signal and we could not identify any resonance structure related to light-induced losses. As an arbitrary choice for our further experiments, we set the wavelength to \SI{2002.00}{nm}. 

All measurements on the intensity dependence are carried out with the molecules held in a single-beam ODT, with a waist of 72, 90, 90, and \SI{120}{\micro m} for the wavelengths 1051, 1064, 1547, and 2002\,nm, respectively.
{For calculating the optical trapping potentials, we assume corresponding values for the dynamic polarizabilities of 805, 784, 557, 510\,a.u. These values are obtained from knowledge of the polarizabilities for both atomic species (see \cite{Dzuba2011dpa, Li2017oto, Ravensbergen2018poa} for Dy and \cite{Safronova2013mwf, Kiruga2025pfh} for K) together with the assumption that the polarizability of a weakly bound molecule can be approximated by the sum of the atomic polarizabilities. The latter turns out to be reasonable as long as accidental resonances of the trap light with optical transitions to excited molecular states, leading to anomalous behavior \cite{Melbourne}, can be avoided.
}

To extract the loss rates for the different cases, we record decay curves for hold times of up to \SI{1}{s}. Assuming one-body effects (processes involving one molecule) to dominate, we fit the different loss curves, except the $2002$-nm case, with a simple exponential decay, $N(t)=N_0\exp{(-\alpha t)}$, where the initial molecule number $N_0$ and the loss rate coefficient $\alpha$ are kept as free parameters. For the $2002$-nm case, the one-body contribution to loss is strongly reduced and two-body contributions (dimer-dimer collisions) to the decay process can no longer be neglected. Consequently, we fit the lifetime data, taking into account both one- and two-body contributions according to the expression \cite{Jag2016lof}
\begin{equation}\label{Eq:12bodyfit}
    N(t)=N_0\,\frac{\exp{(-\alpha t)}}{1+\frac{N_0\,\beta}{\alpha\, \rm{V}_{eff}}[1-\exp{(-\alpha t)}]},
\end{equation}
where $\beta$ is the two-body loss coefficient, and ${V_{\rm eff}}=[(4\pi k_BT)/(m\bar{\omega}^2)]^{3/2}$ is the effective volume of the sample, with $m$ being the mass of the dimer, $T$ the temperature and $\bar{\omega}$ the mean trapping frequency. We will discuss in detail the role of collisional losses in Sec.~\ref{Sec:Collisions}. 

Trap-light-induced losses in Feshbach molecules can essentially be attributed to the coupling between the trapping light and the closed-channel fraction of the weakly bound dimers. The role of the closed-channel fraction has been studied for the case of Li$_2$ molecules \cite{Partridge2005mpo}, for KRb molecules \cite{Chotia2012lld}, and in our group for DyK molecules \cite{Soave2023otf}. To facilitate a comparison between our different measurements, taken at magnetic-field detunings between \SI{-30}{mG} and \SI{-60}{mG}, we 
convert the measured values of the loss rate $\alpha$, which depend on the magnetic field, to the asymptotic loss rate $\alpha_{\rm cc}$ using the relation
\begin{equation}\label{Eq:LossRateMolecules}
\alpha(B)=\alpha_{\rm cc}Z(B),
\end{equation}
where $Z(B)$ is the closed-channel fraction as calculated according to Eq.~(\ref{Eq:ClosedChannelFraction}). The different values for closed-channel fraction for each measurement are reported in Table \ref{tab:Coeff}. 

\begin{figure}[tb]
\centering
\includegraphics[width=1\columnwidth]{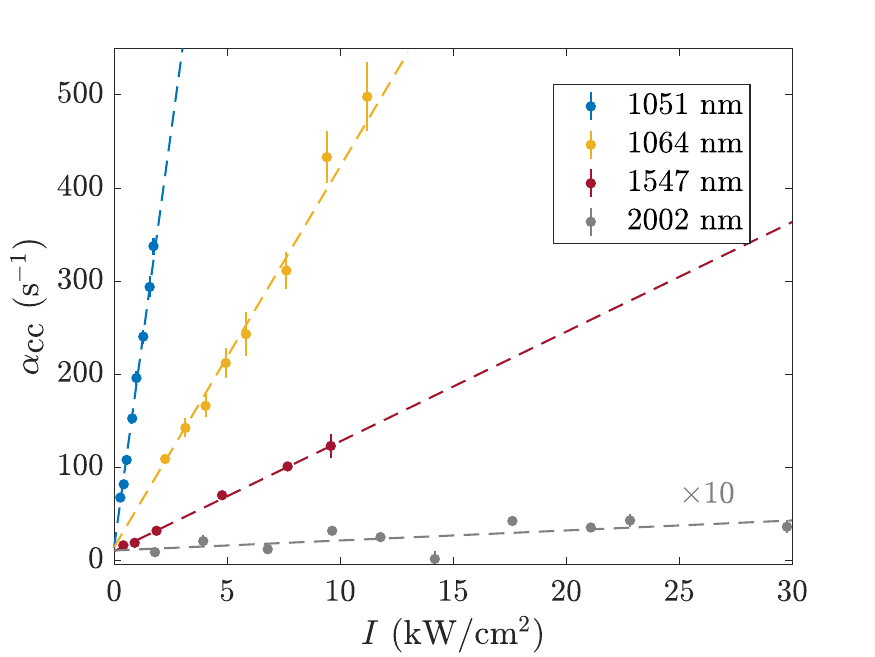}
\caption{Loss rate as a function of the trapping beam peak intensity. We show the closed-channel one-body loss rates measured in the experiments for the \SI{1051.03}{nm}, \SI{1547.12}{nm}, and  \SI{2002.00}{nm} ODTs in blue, red and gray dots, respectively. Note that the data and fit result for the $2002$-nm case have been scaled by a factor of 10 for better visibility. For comparison, we also show the data for a trap at \SI{1064.04}{nm} in yellow, already reported in \cite{Soave2023otf}. The data are shown with 1$\sigma$ error bars derived from the lifetime fits (in some cases smaller than the symbol size).
{For conversion of intensity (in kW/cm$^2$) to optical potential depth (in \SI{}{\micro K}) use the factors 1.76, 1.81, 1.25, and 1.15 for the wavelength 1051\,nm, 1064\,nm, 1547\,nm, and 2002\,nm, respectively.}}
\label{fig:LossRateVsInt}
\end{figure}

In Fig.~\ref{fig:LossRateVsInt}, we show the closed-channel loss rate $\alpha_{\textrm{cc}}$ as function of the light intensity $I$ for the different ODTs, together with a linear fit
\begin{equation}
    \alpha_{\textrm{cc}}=\Gamma_{\textrm{cc}}I+\alpha_0,
    \label{eq:LinFit}
\end{equation}
to each data set. These fits yield a wavelength-specific coefficient $\Gamma_{\rm cc}$, which quantifies the slope for each data set \footnote{Because of the thermal spatial distribution in the trap, the molecules sample a mean intensity, which is typically 25$\,\%$ below the peak intensity. Therefore our values of $\Gamma_{\rm cc}$, which are simply calculated with the peak intensity, underestimate the true values. {In the special case of the 2002-nm trap light, interference effects may lead to locally increased trapping potentials. However, these effects remain small in comparison with the enormous variation of $\Gamma_{\rm cc}$ observed in our experiments.}}. 
The resulting values are summarized in Table~\ref{tab:Coeff}. They show a dramatic variation of $\Gamma_{\textrm{cc}}$ over more than three orders of magnitude. {The linear fit also indicates the presence of residual losses in the limit $I\rightarrow 0$ (parameter $\alpha_0$). Such losses, which cannot be explained in terms of trap-light-induced decay, are likely due to technical imperfections when operating ultra-shallow ODTs and play no role for our present experiments.}
Measurements on light-induced decay have been reported for LiCr Feshbach molecules in Ref.~\cite{Finelli2024ula}, these experiment also demonstrating an order-of-magnitude reduction with increasing wavelength in the infrared. 

\begin{table}[]
\caption{Measured values for the coefficient $\Gamma_{\rm cc}$, which for light-induced losses characterizes proportionality between loss rate and light intensity. The 1051-nm and 1547-nm values are obtained after careful optimization of the wavelength to avoid resonant processes, while the 1064-nm and 2002-nm case are at an arbitrary wavelength (see text for details). We also report the values of the closed-channel fraction $Z(B)$ used to calculate the closed-channel loss rates for each wavelength, according to Eq.~\eqref{Eq:ClosedChannelFraction}.}

    \medskip
    \begin{tabular}{ccc}
     \hline
        \hline
       Wavelength (nm) \qquad&  $Z$ \qquad& $\Gamma_{\rm cc}$ (s$^{-1}$\,cm$^{2}$/kW) \\
       \hline
$1051.03$     & 0.62    &  177(7) \\
$1064.04$ &  0.50 & \   41(2)\\
$1547.12$& 0.58 & 8.4(7) \\
$2002.00$ & 0.58 &   0.11(5)\\
\hline
\hline
    \end{tabular}
    \label{tab:Coeff}
\end{table}

The observed decrease of the light-induced loss rates with larger wavelengths, i.e.\ lower photon energies, can be explained by a reduced density of electronically excited molecular states that can resonantly couple to the Feshbach molecules. Below a certain threshold wavelength, which is given by the energy difference between the lowest molecular state (the rovibronic ground-state of lowest molecular potential), resonant excitation is no longer possible. This wavelength sets a natural scale for the appearance or disappearance of light-induced losses.
For diatomic molecules in ultracold gases, it typically lies in a range between \SI{1.7}{\micro m} and \SI{2.8}{\micro m} \footnote{Examples for the threshold wavelength of bi-alkali molecules are
\SI{1820}{\nano m} for KRb \cite{Ni2008ahp},
\SI{1890}{\nano m} for NaCs \cite{Warner2023ept},
\SI{2000}{\nano m} for RbCs \cite{Takekoshi2014uds}.
For dimers involving other atoms:
\SI{2200}{\nano m} for RbSr \cite{Zuchowski2014gae},
\SI{2660}{\nano m} for LiCr \cite{Finelli2024ula}.
}.

Our results highlight the benefit of laser sources in the wavelength region of \SI{2}{\micro m} for trapping of ultracold molecules. They combine extreme detuning (previously realized in quasi-electrostatic traps based on CO$_2$ laser light at \SI{10.6}{\micro m} \cite{Takekoshi1998ooo, Grimm2000odt}) to avoid any molecular excitations with the practical advantages of modern fiber laser technology.
Currently, the \SI{2}{\micro m} wavelength region is not common in molecular quantum-gas experiments, but it holds great potential for future experiments. 

In view of future improvements, we note that our present setup was not designed to use such a wavelength. In particular, the viewports of the vacuum apparatus are not anti-reflection coated for this wavelength, and we observed reflections of the trap light of up to 30\% when passing through a single viewport. Scanning the laser wavelength, {we found that these reflections caused interference effects such as spatial modulations of the trapping potential.} However, in future experiments, this technical limitation can be overcome by using dedicated optics in an improved setup, which will allow us to take full advantage of  the wavelength region around \SI{2}{\micro m}. 

\section{Collisional Losses and Pauli Suppression}
\label{Sec:Collisions}

Having understood the role of losses induced by the trap light, we are now in a position to thoroughly investigate the contribution of inelastic collisions to the decay of the molecular sample. In particular, it is well known that, in the universal regime close to the resonance center, collisional losses are reduced by Pauli suppression. This has been observed in samples of weakly bound bosonic molecules composed of fermionic atoms ($^6$Li \cite{Cubizolles2003pol, Jochim2003pgo}, $^{40}$K \cite{Regal2004lom} and $^6$Li$^{40}$K \cite{Jag2016lof}).

In order to observe collisional processes in the DyK molecular sample, it is essential to minimize light-induced decay while achieving a high collisional rate. Besides optimization of the trap light wavelength, this requires an ODT that provides tight confinement at a relatively low central intensity. This means that tightly focused laser beams need to be applied. Because of the technical limitations in our present set-up, this is not possible with the 2002-nm light, and we therefore conducted this measurement with the 1547-nm laser light. 

The trap consists of a crossed-beam ODT at \SI{1547}{nm}. The main trapping beam propagates along the horizontal plane and is focused on the molecules with a waist of \SI{25}{\micro m}. For additional confinement in the axial direction of the tight trap, we add a vertical beam, with a waist of \SI{120}{\micro\meter}. A magnetic gradient is applied to levitate the molecules. In this ODT, the geometrically averaged trap frequency is $\bar\omega=2\pi\times \SI{32}{Hz}$. With this configuration, we obtain between $5500$ and $7000$ molecules, at a temperature of around $\SI{90}{nK}$, corresponding to peak densities up to \SI{7.6e11}{cm^{-3}}, entering the regime where collisional losses play an important role. 
The power used for the horizontal (vertical) beam is \SI{4.6}{mW} (\SI{51}{mW}). The total light intensity is $\SI{0.71}{kW/cm^{2}}$. Based on the results of Sec.~\ref{subSec:LightIntensity}, we can expect a closed-channel loss rate of $\alpha_{\rm cc} = \SI{6.0(7)}{s^{-1}}$.

\begin{figure}[tb]
\centering
\includegraphics[width=1\columnwidth]{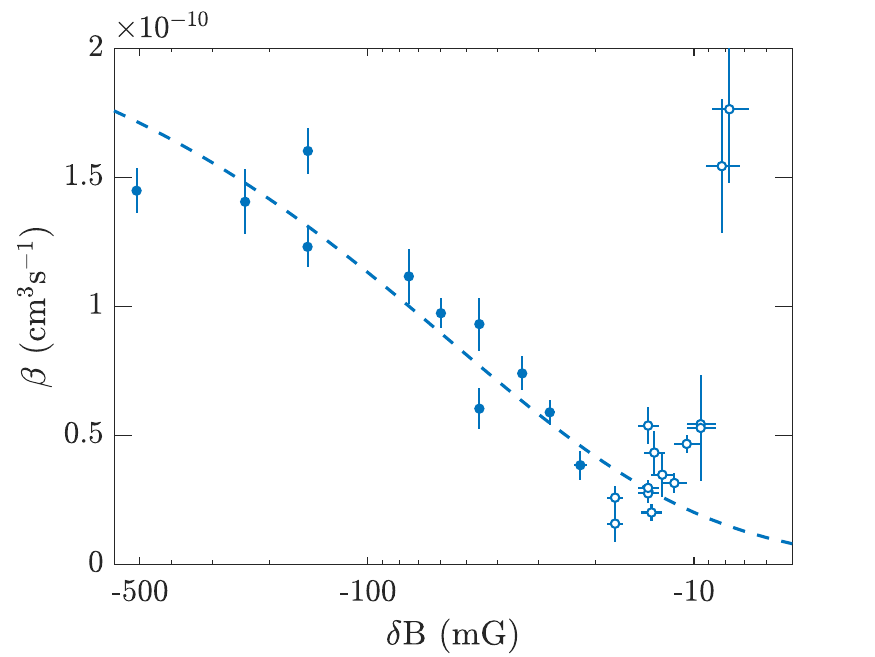}
\caption{Two-body loss rate coefficient versus magnetic detuning from resonance. The dashed line represents a one-parameter fit of Eq.~\eqref{eq:FitPauli} to the data points for detunings $\delta B<-\SI{20}{mG}$ (filled symbols). Additional measurements (open symbols) closer to resonance, presumably beyond the limitations of our simple model, have not been taken into account for the fit. Vertical error bars indicate $1\sigma$ errors, derived from fits to the decay curves. The horizontal error bars represent the 2-mG peak-to-peak magnetic field noise.}
\label{Fig:Pauli}
\end{figure}

In order to study the Pauli suppression of inelastic collisions expected close to resonance, we measure the two-body coefficient $\beta$ for different detunings from the Feshbach resonance. We extract values for $\beta$ by analyzing the decay of the sample at different magnetic field strengths and fitting Eq.~(\ref{Eq:12bodyfit}) to the data, with $N_0$, $\alpha$, and $\beta$ as free parameters. 
The values obtained in this way for $\alpha$ (not shown) follow the general behavior according to Eq.~(\ref{Eq:LossRateMolecules}) and yield $\alpha_{\rm cc} = \SI{6.2(2)}{s^{-1}}$, which is fully consistent with the results of Sec.~\ref{subSec:LightIntensity}, and follows the behavior observed in Ref.~\cite{Soave2023otf}. Our experimental values for the magnetic-field dependent two-body coefficient $\beta$ are displayed in Fig.~\ref{Fig:Pauli}. We indeed observe a clear reduction of $\beta$ for magnetic fields approaching the resonance center, which we interpret as the expected suppression effect.

In Ref.~\cite{Jag2016lof}, collisional relaxation of weakly bound dimers composed of fermionic atoms has been modeled, considering the situation near a narrow Feshbach resonance. In this work, DyK molecules have served as a specific example. According to the model, collisional relaxation takes place in three different channels, the two atom-dimer channels (atoms of Dy or K with closed-channel DyK dimers) and the dimer-dimer channel. Consequently, the total collisional relaxation rate is obtained as a linear combination of the relaxation rates for each channel weighted with a specific suppression function that contains the Pauli blocking effect.

\begin{figure}[tb]
    \centering
    \includegraphics[width=1\linewidth]{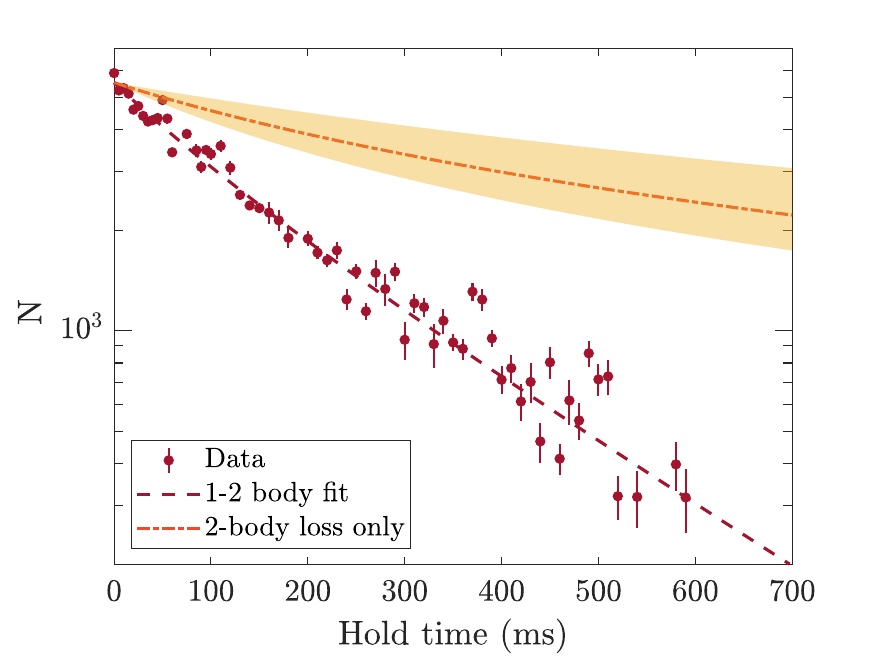}
    \caption{Collisional decay and the role of light-induced decay, observed in a 1547-nm trap. The red dots show the number of molecules measured after a variable hold time. The red dashed line represents a fit according to Eq.~(\ref{Eq:12bodyfit}), which yields the parameter values $\alpha = 4.2(4)\,{\rm s}^{-1}$, $\beta=1.5(0.7)\times 10^{-11}\,{\rm cm^3\,s^{-1}}$ and $N_0 = 5530(130)$.
    The orange dash-dotted line shows a hypothetical loss curve with the same two-body loss in the complete absence of one-body losses ($\alpha\rightarrow 0$). The shaded region reflects the 1-$\sigma$ uncertainty in $\beta$.}
    \label{fig:LongestLife}
\end{figure}

For magnetic detunings outside of the universal range ($a\lesssim R^*$ corresponding to $|\delta B| \gtrsim 4\delta B^*$), the model can be further simplified {\cite{Jag2016lof}}. Here the dimer-dimer channel dominates the relaxation and the contribution of the two atom-dimer channels can be neglected. The suppression function for the dimer-dimer channel simplifies to the squared closed-channel fraction, $Z^2(B)$, and we can approximate the relaxation rate coefficient by
\begin{equation}
    \beta(B)=\beta_0\,Z^2(B),
    \label{eq:FitPauli}
\end{equation}
with $\beta_0$ as the only free parameter. A corresponding fit to the experimental data, shown by the dashed line reported in Fig.~\ref{Fig:Pauli}, describes well the data for $\delta B< -\SI{20}{mG}$, a range over which the closed-channel character of the molecules is expected to prevail. From the fit we obtain the closed-channel dimer-dimer asymptotic rate coefficient $\beta_0=\SI{2.3(1)e-10}{cm^3s^{-1}}$. This value is comparable with predictions from a quantum Langevin model calculated for similar systems \cite{Gao2010umf, Julienne2011uuc, Quenemer2011uiu}. The minimum value $\beta\approx\SI{2e-11}{cm^3s^{-1}}$ is obtained around  the detuning $\delta B=-\SI{17}{mG}$ (see also Fig.\,\ref{fig:LongestLife}) and shows a reduction of $\beta$ by about an order of magnitude, compared to the asymptotic value.

Closer to resonance, a further decrease of $\beta$ may be expected as a consequence of the increasing role of the fermionic nature of the dimer's constituents \cite{Jag2016lof}. However,  for $\delta B \gtrsim-$\SI{13}{mG}, we observe a rapid increase of the two-body coefficient. We attribute this to a combination of different effects leading to a dissociation of the very weakly bound dimers. One contribution is caused by the magnetic field gradient, which makes the exact resonance position dependent on the vertical position. At a detuning of \SI{-10}{mG} the resonance pole is located only few tens of \SI{}{\micro m} above the center of the molecular cloud, which is comparable with the vertical extension of the cloud. Moreover, as a rather technical issue, the transfer of the molecules between different traps can excite a weak dipole mode in the vertical direction, bringing the molecules even closer to the resonance pole. Additionally, in this region of the magnetic field, the binding energy of the molecules is only a few hundred of \SI{}{nK}, less than an order of magnitude above the thermal kinetic energy of the molecules. In this regime, endoergic collisions can lead to dissociation of the weakly bound dimers \cite{Jochim2003pgo, Chin2004tea}. 
{The resulting atoms do not fulfill the magnetic levitation condition and cannot be kept in the trap. In this way, the inherent Stern-Gerlach selection effect~\cite{Soave2023otf}, which is not present for homonuclear dimers, steadily purifies the molecular sample (see also Sec.~\ref{Sec:SamplePrep}).}
Understanding the particular role and interplay of these dissociation effects will require more investigations, e.g.\ in optical trapping schemes combined with homogeneous magnetic fields. Closer to resonance there is potential for further enhancing the fermionic suppression effect. 

In Fig.\,\ref{fig:LongestLife}, we show the observed decay of the molecular sample at the detuning $\delta B=-\SI{17}{mG}$, where we have identified minimal collisional losses (lowest value of $\beta$ in Fig.~\ref{Fig:Pauli}), together with the best fit according to Eq.~(\ref{Eq:12bodyfit}). As a guide for future experiments, we consider the hypothetical case where light-induced losses are completely absent and only two-body collisional losses remain. Thus setting $\alpha \rightarrow 0$ and using the fitted parameter values for $N_0$ and $\beta$, we obtain the decay curve expected in the absence of light-induced losses, shown as the orange line in the figure. This analysis demonstrates that, in a trapping regime free of light-induced losses, substantially longer decay times are achievable than in our present experiments.

Finally, it is very instructive to compare the elastic collision rate in the trapped dimer sample with the {minimum two-body loss rate under the best conditions we could so far achieve ($\delta B = -17\, {\rm mG}$). We further assume the complete suppression of trap-light induced one-body losses ($\alpha = 0$) in an ideal ODT.} The elastic scattering rate can be calculated as 
$\gamma_{\rm el} = \bar{n} \sigma \bar{v}_{\rm rel}$, 
with the mean number density 
$\bar{n} = N/V_{\rm eff} $ the mean relative velocity $\bar{v}_{\rm rel} = (16 k_B T/(\pi m))^{1/2}$,
and the scattering cross section $\sigma = 8\pi a_{\rm mol}^2$, where $a_{\rm mol} = 0.77\,a$ represents the dimer-dimer scattering length \cite{Petrov2005dmi, Marcelis2008cpo}. For the experimental parameters of Fig.~\ref{fig:LongestLife} ($N= 5500$, $\bar{\omega} = 2\pi \times 32$\,Hz, $T = 90\,$nK), we obtain $\gamma_{\rm el} = 48\,{\rm s}^{-1}$. For the loss rate, we calculate $\gamma_{\rm loss} = \bar{n} \beta = 3.4\,{\rm s}^{-1}$.
The ratio $\gamma_{\rm el}/\gamma_{\rm loss} \approx 14$ tells us that elastic collisions dominate over inelastic ones, but only by about one order of magnitude. 
Consequently, our attempts to implement evaporative cooling in the present set-up were not successful.

In view of prospective experiments involving quantum-degenerate molecular samples, it is an interesting question whether conditions can be reached for efficient evaporative cooling, which typically requires a good-to-bad collision ratio $\gamma_{\rm el}/\gamma_{\rm loss} \gg 100$  \cite{ketterle1996eco}. If we assume a reduction of the magnetic resonance detuning from the present optimum $\delta B = -17\,$mG by just a factor of two, which corresponds to an increase of the scattering length $a$ by a factor of two, this would increase the elastic scattering rate $\gamma_{\rm el} \propto a^2$ by a factor of four. At the same time, according to the scaling predicted for the Dy-K mass ratio \cite{Petrov2005dmi} $\gamma_{\rm loss} \propto a^{1.8}$, the inelastic rate would drop by a factor of about $3.5$. These two effects together would boost the good-to-bad collision ratio by a factor of $14$ to $\gamma_{\rm el}/\gamma_{\rm loss} \approx 200$, which seems promising for evaporative cooling. To take full advantage of this in an experiment, a highly stable, homogeneous magnetic field would be necessary along with an optical trapping scheme that does not require magnetic levitation.

\section{Conclusions and outlook}
\label{Sec:Conclusions}

We have studied the decay of a dense, ultracold sample of weakly bound DyK dimers stored in an optical dipole trap that operates with far red-detuned laser light. We consider the case of bosonic dimers composed of the fermionic isotopes $^{161}$Dy and $^{40}$K, which is of particular interest for experiments related to pairing and superfluidity in fermionic systems with mass imbalance. We have identified trap-light-induced and collisional processes as the two main sources of losses, and we have demonstrated ways to reduce these losses substantially.

Light-induced losses have been investigated in four selected spectral windows in a wide near-infrared wavelength range between 1051 and 2002\,nm. We have measured corresponding loss rate coefficients, which we found to vary over four orders of magnitude with a clear general trend of strong decrease with larger wavelengths. In addition to that, we have observed resonant loss features, pointing to bound-bound electronic transitions, and plateaus of minimal losses, which are of particular interest for specific applications. Background losses seem to be present everywhere in the spectral range investigated, as our sample spectra suggest. 
Qualitatively, this behavior can be explained by the very complex and dense spectrum of molecular states in a dimer with complex electronic structure, which features a multitude of different potential curves. With increasing wavelength, i.e.\ lower photon energy, less molecular potentials become accessible and the resonance density decreases until resonant excitation is no longer possible beyond a certain point (wavelengths typically $\gtrsim$\,2\,\unit{\micro m}). A comprehensive interpretation of all the features observed in our experiments would require detailed knowledge of the spectrum of molecular states, which is currently not available for the DyK system.

While light-induced losses constitute a detrimental effect for the specific application discussed in our present work, they more generally offer a sensitive probe for pair correlations and related interaction phenomena in resonant quantum gases~\cite{Partridge2005mpo, Liu2021oot, Jaeger2024ppm, Journeaux2026tbc}. When harnessed as an experimental diagnostics tool, these losses can provide unique insights into the role of closed-channel dimers near narrow Feshbach resonances and the system's rapid non-equilibrium dynamics. Such potential applications motivate further investigations of light-induced effects in near-resonant systems, including the Dy-K mixture studied here and other single- or mixed-species quantum gases.

Collisional losses {in our system} result from inelastic dimer-dimer collisions. {For their experimental observation}, it was necessary to minimize light-induced decay by carefully choosing the wavelength of the ODT (1547.21\,nm regarding the available laser sources). For dimer binding energies exceeding $\sim h \times 1\,$MHz, we measured collisional decay rates typical for cold dimers in high-lying vibrational states. For a much lower binding energy of about 20\,kHz (closer to the center of the Feshbach resonance employed), we observed a clear suppression of inelastic collisional losses by roughly one order of magnitude. We interpret this observation in terms of the famous Pauli suppression effect that stabilizes weakly bound dimer samples and fermionic mixtures near Feshbach resonances and enables experiments in the resonance regime. 
In principle, much stronger loss suppression can be expected closer to the resonance, but  our  present experiments were technically limited by magnetic field inhomogeneities caused by the magnetic levitation gradient that was needed to keep the cold sample in the trap.

In view of future experiments, our work conveys important messages. Optically trapped molecules, in particular those with complex electronic structure, will be in general much more susceptible to detrimental light-induced processes than we are used from the bi-alkali dimers commonly used in many experiments. Thus, experiments will require a careful choice of the particular trap-light wavelength. The wavelength region around 2\,\unit{\micro m}, where high-power fiber laser sources are now available, appears to be particularly promising for eliminating light-induced losses. For our special application related to mass-imbalanced Fermi gases, we could confirm the presence of Pauli suppression of collisional losses near a Feshbach resonance. However, to take full advantage of this effect in the DyK system, a trap set-up is needed that allows one to work in a precisely controllable homogeneous magnetic field. These insights will be essential for planning a second-generation set up.

\section*{Acknowledgments}

We thank M.\,Zaccanti and the members of the Li-Cr team at LENS, as well as D.\,Petrov for  stimulating discussions. We also wish to thank all members of the ultracold groups in Innsbruck for a stimulating environment and many insights related to the broad field of quantum-gas mixtures. The project has received funding from the European Research Council (ERC) under the European Union’s Horizon 2020 research and innovation programme (Grant Agreement No.~101020438 - SuperCoolMix). We further acknowledge support by the Austrian Science Fund (FWF) within the Doktoratskolleg ALM (W1259-N27). Z.-X.\,Y.~acknowledges support by the National Natural Science Foundation of China (No.~U23A6004, No.~12504336) and the Fundamental Research Program of Shanxi Province (No.~202403021221001).

\bigskip
{
{\em Data availability} -- The data that support the findings of this article are openly available \cite{repository}.
}


\begin{thebibliography}{101}%
\makeatletter
\providecommand \@ifxundefined [1]{%
 \@ifx{#1\undefined}
}%
\providecommand \@ifnum [1]{%
 \ifnum #1\expandafter \@firstoftwo
 \else \expandafter \@secondoftwo
 \fi
}%
\providecommand \@ifx [1]{%
 \ifx #1\expandafter \@firstoftwo
 \else \expandafter \@secondoftwo
 \fi
}%
\providecommand \natexlab [1]{#1}%
\providecommand \enquote  [1]{``#1''}%
\providecommand \bibnamefont  [1]{#1}%
\providecommand \bibfnamefont [1]{#1}%
\providecommand \citenamefont [1]{#1}%
\providecommand \href@noop [0]{\@secondoftwo}%
\providecommand \href [0]{\begingroup \@sanitize@url \@href}%
\providecommand \@href[1]{\@@startlink{#1}\@@href}%
\providecommand \@@href[1]{\endgroup#1\@@endlink}%
\providecommand \@sanitize@url [0]{\catcode `\\12\catcode `\$12\catcode
  `\&12\catcode `\#12\catcode `\^12\catcode `\_12\catcode `\%12\relax}%
\providecommand \@@startlink[1]{}%
\providecommand \@@endlink[0]{}%
\providecommand \url  [0]{\begingroup\@sanitize@url \@url }%
\providecommand \@url [1]{\endgroup\@href {#1}{\urlprefix }}%
\providecommand \urlprefix  [0]{URL }%
\providecommand \Eprint [0]{\href }%
\providecommand \doibase [0]{https://doi.org/}%
\providecommand \selectlanguage [0]{\@gobble}%
\providecommand \bibinfo  [0]{\@secondoftwo}%
\providecommand \bibfield  [0]{\@secondoftwo}%
\providecommand \translation [1]{[#1]}%
\providecommand \BibitemOpen [0]{}%
\providecommand \bibitemStop [0]{}%
\providecommand \bibitemNoStop [0]{.\EOS\space}%
\providecommand \EOS [0]{\spacefactor3000\relax}%
\providecommand \BibitemShut  [1]{\csname bibitem#1\endcsname}%
\let\auto@bib@innerbib\@empty
\bibitem [{\citenamefont {Chu}\ \emph {et~al.}(1986)\citenamefont {Chu},
  \citenamefont {Bjorkholm}, \citenamefont {Ashkin},\ and\ \citenamefont
  {Cable}}]{Chu1986eoo}%
  \BibitemOpen
  \bibfield  {author} {\bibinfo {author} {\bibfnamefont {S.}~\bibnamefont
  {Chu}}, \bibinfo {author} {\bibfnamefont {J.~E.}\ \bibnamefont {Bjorkholm}},
  \bibinfo {author} {\bibfnamefont {A.}~\bibnamefont {Ashkin}},\ and\ \bibinfo
  {author} {\bibfnamefont {A.}~\bibnamefont {Cable}},\ }\bibfield  {title}
  {\bibinfo {title} {Experimental observation of optically trapped atoms},\
  }\href {https://doi.org/10.1103/PhysRevLett.57.314} {\bibfield  {journal}
  {\bibinfo  {journal} {Phys. Rev. Lett.}\ }\textbf {\bibinfo {volume} {57}},\
  \bibinfo {pages} {314} (\bibinfo {year} {1986})}\BibitemShut {NoStop}%
\bibitem [{\citenamefont {Grimm}\ \emph {et~al.}(2000)\citenamefont {Grimm},
  \citenamefont {Weidem\"uller},\ and\ \citenamefont
  {Ovchinnikov}}]{Grimm2000odt}%
  \BibitemOpen
  \bibfield  {author} {\bibinfo {author} {\bibfnamefont {R.}~\bibnamefont
  {Grimm}}, \bibinfo {author} {\bibfnamefont {M.}~\bibnamefont
  {Weidem\"uller}},\ and\ \bibinfo {author} {\bibfnamefont {Y.~B.}\
  \bibnamefont {Ovchinnikov}},\ }\bibfield  {title} {\bibinfo {title} {Optical
  dipole traps for neutral atoms},\ }\href
  {https://doi.org/10.1016/S1049-250X(08)60186-X} {\bibfield  {journal}
  {\bibinfo  {journal} {Adv. At. Mol. Opt. Phys.}\ }\textbf {\bibinfo {volume}
  {42}},\ \bibinfo {pages} {95} (\bibinfo {year} {2000})}\BibitemShut {NoStop}%
\bibitem [{\citenamefont {Wolswijk}\ \emph {et~al.}(2025)\citenamefont
  {Wolswijk}, \citenamefont {Cavicchioli}, \citenamefont {Vinelli},
  \citenamefont {Chiarotti}, \citenamefont {Donati}, \citenamefont {Fernandez},
  \citenamefont {Rajkov}, \citenamefont {Mancini}, \citenamefont {Vezio},
  \citenamefont {Zhou}, \citenamefont {Pace}, \citenamefont {Mazzinghi},
  \citenamefont {Antolini}, \citenamefont {Salvi},\ and\ \citenamefont
  {Gavryusev}}]{Wolswijk2025tma}%
  \BibitemOpen
  \bibfield  {author} {\bibinfo {author} {\bibfnamefont {L.}~\bibnamefont
  {Wolswijk}}, \bibinfo {author} {\bibfnamefont {L.}~\bibnamefont
  {Cavicchioli}}, \bibinfo {author} {\bibfnamefont {G.}~\bibnamefont
  {Vinelli}}, \bibinfo {author} {\bibfnamefont {M.}~\bibnamefont {Chiarotti}},
  \bibinfo {author} {\bibfnamefont {L.}~\bibnamefont {Donati}}, \bibinfo
  {author} {\bibfnamefont {M.~F.}\ \bibnamefont {Fernandez}}, \bibinfo {author}
  {\bibfnamefont {D.~H.}\ \bibnamefont {Rajkov}}, \bibinfo {author}
  {\bibfnamefont {C.}~\bibnamefont {Mancini}}, \bibinfo {author} {\bibfnamefont
  {P.}~\bibnamefont {Vezio}}, \bibinfo {author} {\bibfnamefont
  {T.}~\bibnamefont {Zhou}}, \bibinfo {author} {\bibfnamefont {G.~D.}\
  \bibnamefont {Pace}}, \bibinfo {author} {\bibfnamefont {C.}~\bibnamefont
  {Mazzinghi}}, \bibinfo {author} {\bibfnamefont {N.}~\bibnamefont {Antolini}},
  \bibinfo {author} {\bibfnamefont {L.}~\bibnamefont {Salvi}},\ and\ \bibinfo
  {author} {\bibfnamefont {V.}~\bibnamefont {Gavryusev}},\ }\href
  {https://arxiv.org/abs/2510.20790} {\bibinfo {title} {Trapping, manipulating
  and probing ultracold atoms: a quantum technologies tutorial}} (\bibinfo
  {year} {2025}),\ \Eprint {https://arxiv.org/abs/2510.20790} {arXiv:2510.20790
  [cond-mat.quant-gas]} \BibitemShut {NoStop}%
\bibitem [{\citenamefont {Ludlow}\ \emph {et~al.}(2015)\citenamefont {Ludlow},
  \citenamefont {Boyd}, \citenamefont {Ye}, \citenamefont {Peik},\ and\
  \citenamefont {Schmidt}}]{Ludlow2015oac}%
  \BibitemOpen
  \bibfield  {author} {\bibinfo {author} {\bibfnamefont {A.~D.}\ \bibnamefont
  {Ludlow}}, \bibinfo {author} {\bibfnamefont {M.~M.}\ \bibnamefont {Boyd}},
  \bibinfo {author} {\bibfnamefont {J.}~\bibnamefont {Ye}}, \bibinfo {author}
  {\bibfnamefont {E.}~\bibnamefont {Peik}},\ and\ \bibinfo {author}
  {\bibfnamefont {P.~O.}\ \bibnamefont {Schmidt}},\ }\bibfield  {title}
  {\bibinfo {title} {Optical atomic clocks},\ }\href
  {https://doi.org/10.1103/RevModPhys.87.637} {\bibfield  {journal} {\bibinfo
  {journal} {Rev. Mod. Phys.}\ }\textbf {\bibinfo {volume} {87}},\ \bibinfo
  {pages} {637} (\bibinfo {year} {2015})}\BibitemShut {NoStop}%
\bibitem [{\citenamefont {Yang}\ \emph {et~al.}(2025)\citenamefont {Yang},
  \citenamefont {Miklos}, \citenamefont {Tso}, \citenamefont {Kraus},
  \citenamefont {Hur},\ and\ \citenamefont {Ye}}]{Yang2025cbb}%
  \BibitemOpen
  \bibfield  {author} {\bibinfo {author} {\bibfnamefont {Y.~A.}\ \bibnamefont
  {Yang}}, \bibinfo {author} {\bibfnamefont {M.}~\bibnamefont {Miklos}},
  \bibinfo {author} {\bibfnamefont {Y.~M.}\ \bibnamefont {Tso}}, \bibinfo
  {author} {\bibfnamefont {S.}~\bibnamefont {Kraus}}, \bibinfo {author}
  {\bibfnamefont {J.}~\bibnamefont {Hur}},\ and\ \bibinfo {author}
  {\bibfnamefont {J.}~\bibnamefont {Ye}},\ }\bibfield  {title} {\bibinfo
  {title} {Clock precision beyond the standard quantum limit at
  ${10}^{\ensuremath{-}18}$ level},\ }\href {https://doi.org/10.1103/6v93-whwq}
  {\bibfield  {journal} {\bibinfo  {journal} {Phys. Rev. Lett.}\ }\textbf
  {\bibinfo {volume} {135}},\ \bibinfo {pages} {193202} (\bibinfo {year}
  {2025})}\BibitemShut {NoStop}%
\bibitem [{\citenamefont {Zheng}\ \emph {et~al.}(2022)\citenamefont {Zheng},
  \citenamefont {Yang}, \citenamefont {Wang}, \citenamefont {Singh},
  \citenamefont {Xiong}, \citenamefont {Xia},\ and\ \citenamefont
  {Lu}}]{Zheng2022mot}%
  \BibitemOpen
  \bibfield  {author} {\bibinfo {author} {\bibfnamefont {T.~A.}\ \bibnamefont
  {Zheng}}, \bibinfo {author} {\bibfnamefont {Y.~A.}\ \bibnamefont {Yang}},
  \bibinfo {author} {\bibfnamefont {S.-Z.}\ \bibnamefont {Wang}}, \bibinfo
  {author} {\bibfnamefont {J.~T.}\ \bibnamefont {Singh}}, \bibinfo {author}
  {\bibfnamefont {Z.-X.}\ \bibnamefont {Xiong}}, \bibinfo {author}
  {\bibfnamefont {T.}~\bibnamefont {Xia}},\ and\ \bibinfo {author}
  {\bibfnamefont {Z.-T.}\ \bibnamefont {Lu}},\ }\bibfield  {title} {\bibinfo
  {title} {Measurement of the electric dipole moment of $^{171}\mathrm{Yb}$
  atoms in an optical dipole trap},\ }\href
  {https://doi.org/10.1103/PhysRevLett.129.083001} {\bibfield  {journal}
  {\bibinfo  {journal} {Phys. Rev. Lett.}\ }\textbf {\bibinfo {volume} {129}},\
  \bibinfo {pages} {083001} (\bibinfo {year} {2022})}\BibitemShut {NoStop}%
\bibitem [{\citenamefont {Thekkeppatt}\ \emph {et~al.}(2025)\citenamefont
  {Thekkeppatt}, \citenamefont {Digvijay}, \citenamefont {Urech}, \citenamefont
  {Schreck},\ and\ \citenamefont {van Druten}}]{Thekkeppatt2025mot}%
  \BibitemOpen
  \bibfield  {author} {\bibinfo {author} {\bibfnamefont {P.}~\bibnamefont
  {Thekkeppatt}}, \bibinfo {author} {\bibnamefont {Digvijay}}, \bibinfo
  {author} {\bibfnamefont {A.}~\bibnamefont {Urech}}, \bibinfo {author}
  {\bibfnamefont {F.}~\bibnamefont {Schreck}},\ and\ \bibinfo {author}
  {\bibfnamefont {K.}~\bibnamefont {van Druten}},\ }\bibfield  {title}
  {\bibinfo {title} {Measurement of the $g$ factor of ground-state
  $^{87}\mathrm{Sr}$ at the parts-per-million level using co-trapped ultracold
  atoms},\ }\href {https://doi.org/10.1103/cjks-9hlp} {\bibfield  {journal}
  {\bibinfo  {journal} {Phys. Rev. Lett.}\ }\textbf {\bibinfo {volume} {135}},\
  \bibinfo {pages} {193001} (\bibinfo {year} {2025})}\BibitemShut {NoStop}%
\bibitem [{\citenamefont {Ravensbergen}\ \emph
  {et~al.}(2018{\natexlab{a}})\citenamefont {Ravensbergen}, \citenamefont
  {Corre}, \citenamefont {Soave}, \citenamefont {Kreyer}, \citenamefont
  {Tzanova}, \citenamefont {Kiri\-lov},\ and\ \citenamefont
  {Grimm}}]{Ravensbergen2018ado}%
  \BibitemOpen
  \bibfield  {author} {\bibinfo {author} {\bibfnamefont {C.}~\bibnamefont
  {Ravensbergen}}, \bibinfo {author} {\bibfnamefont {V.}~\bibnamefont {Corre}},
  \bibinfo {author} {\bibfnamefont {E.}~\bibnamefont {Soave}}, \bibinfo
  {author} {\bibfnamefont {M.}~\bibnamefont {Kreyer}}, \bibinfo {author}
  {\bibfnamefont {S.}~\bibnamefont {Tzanova}}, \bibinfo {author} {\bibfnamefont
  {E.}~\bibnamefont {Kiri\-lov}},\ and\ \bibinfo {author} {\bibfnamefont
  {R.}~\bibnamefont {Grimm}},\ }\bibfield  {title} {\bibinfo {title} {{Accurate
  Determination of the Dynamical Polarizability of Dysprosium}},\ }\href
  {https://doi.org/10.1103/PhysRevLett.120.223001} {\bibfield  {journal}
  {\bibinfo  {journal} {Phys. Rev. Lett.}\ }\textbf {\bibinfo {volume} {120}},\
  \bibinfo {pages} {223001} (\bibinfo {year} {2018}{\natexlab{a}})}\BibitemShut
  {NoStop}%
\bibitem [{\citenamefont {Browaeys}\ and\ \citenamefont
  {Lahaye}(2020)}]{Browaeys2020mbp}%
  \BibitemOpen
  \bibfield  {author} {\bibinfo {author} {\bibfnamefont {A.}~\bibnamefont
  {Browaeys}}\ and\ \bibinfo {author} {\bibfnamefont {T.}~\bibnamefont
  {Lahaye}},\ }\bibfield  {title} {\bibinfo {title} {{Many-body physics with
  individually controlled Rydberg atoms}},\ }\href
  {https://doi.org/10.1038/s41567-019-0733-z} {\bibfield  {journal} {\bibinfo
  {journal} {Nat. Phys.}\ }\textbf {\bibinfo {volume} {16}},\ \bibinfo {pages}
  {132} (\bibinfo {year} {2020})}\BibitemShut {NoStop}%
\bibitem [{\citenamefont {Kaufman}\ and\ \citenamefont
  {Ni}(2021)}]{Kaufman2021qsw}%
  \BibitemOpen
  \bibfield  {author} {\bibinfo {author} {\bibfnamefont {A.~M.}\ \bibnamefont
  {Kaufman}}\ and\ \bibinfo {author} {\bibfnamefont {K.-K.}\ \bibnamefont
  {Ni}},\ }\bibfield  {title} {\bibinfo {title} {{Quantum science with optical
  tweezer arrays of ultracold atoms and molecules}},\ }\href
  {https://doi.org/10.1038/s41567-021-01357-2} {\bibfield  {journal} {\bibinfo
  {journal} {Nat. Phys.}\ }\textbf {\bibinfo {volume} {17}},\ \bibinfo {pages}
  {1324} (\bibinfo {year} {2021})}\BibitemShut {NoStop}%
\bibitem [{\citenamefont {Anand}\ \emph {et~al.}(2024)\citenamefont {Anand},
  \citenamefont {Bradley}, \citenamefont {White}, \citenamefont {Ramesh},
  \citenamefont {Singh},\ and\ \citenamefont {Bernien}}]{Anand2024ads}%
  \BibitemOpen
  \bibfield  {author} {\bibinfo {author} {\bibfnamefont {S.}~\bibnamefont
  {Anand}}, \bibinfo {author} {\bibfnamefont {C.~E.}\ \bibnamefont {Bradley}},
  \bibinfo {author} {\bibfnamefont {R.}~\bibnamefont {White}}, \bibinfo
  {author} {\bibfnamefont {V.}~\bibnamefont {Ramesh}}, \bibinfo {author}
  {\bibfnamefont {K.}~\bibnamefont {Singh}},\ and\ \bibinfo {author}
  {\bibfnamefont {H.}~\bibnamefont {Bernien}},\ }\bibfield  {title} {\bibinfo
  {title} {{A dual-species Rydberg array}},\ }\href
  {https://doi.org/10.1038/s41567-024-02638-2} {\bibfield  {journal} {\bibinfo
  {journal} {Nat. Phys.}\ }\textbf {\bibinfo {volume} {20}},\ \bibinfo {pages}
  {1744} (\bibinfo {year} {2024})}\BibitemShut {NoStop}%
\bibitem [{\citenamefont {Gr\"un}\ \emph {et~al.}(2024)\citenamefont {Gr\"un},
  \citenamefont {White}, \citenamefont {Ortu}, \citenamefont {Di~Carli},
  \citenamefont {Edri}, \citenamefont {Lepers}, \citenamefont {Mark},\ and\
  \citenamefont {Ferlaino}}]{Gruen2024ota}%
  \BibitemOpen
  \bibfield  {author} {\bibinfo {author} {\bibfnamefont {D.~S.}\ \bibnamefont
  {Gr\"un}}, \bibinfo {author} {\bibfnamefont {S.~J.~M.}\ \bibnamefont
  {White}}, \bibinfo {author} {\bibfnamefont {A.}~\bibnamefont {Ortu}},
  \bibinfo {author} {\bibfnamefont {A.}~\bibnamefont {Di~Carli}}, \bibinfo
  {author} {\bibfnamefont {H.}~\bibnamefont {Edri}}, \bibinfo {author}
  {\bibfnamefont {M.}~\bibnamefont {Lepers}}, \bibinfo {author} {\bibfnamefont
  {M.~J.}\ \bibnamefont {Mark}},\ and\ \bibinfo {author} {\bibfnamefont
  {F.}~\bibnamefont {Ferlaino}},\ }\bibfield  {title} {\bibinfo {title}
  {{Optical Tweezer Arrays of Erbium Atoms}},\ }\href
  {https://doi.org/10.1103/PhysRevLett.133.223402} {\bibfield  {journal}
  {\bibinfo  {journal} {Phys. Rev. Lett.}\ }\textbf {\bibinfo {volume} {133}},\
  \bibinfo {pages} {223402} (\bibinfo {year} {2024})}\BibitemShut {NoStop}%
\bibitem [{\citenamefont {Inguscio}\ \emph
  {et~al.}({\natexlab{a}})\citenamefont {Inguscio}, \citenamefont {Stringari},\
  and\ \citenamefont {Wieman}}]{Varenna1998book}%
  \BibitemOpen
  \bibinfo {editor} {\bibfnamefont {M.}~\bibnamefont {Inguscio}}, \bibinfo
  {editor} {\bibfnamefont {S.}~\bibnamefont {Stringari}},\ and\ \bibinfo
  {editor} {\bibfnamefont {C.}~\bibnamefont {Wieman}},\ eds.,\ \href@noop {}
  {\emph {\bibinfo {title} {{Bose-Einstein Condensation in Atomic Gases}}}}\
  (\bibinfo  {publisher} {North Holland, Amsterdam, 1999})\ \bibinfo {note}
  {{P}roceedings of the International School of Physics ``Enrico Fermi'',
  Course CXL, Varenna, 7-17 July 1998}\BibitemShut {NoStop}%
\bibitem [{\citenamefont {Inguscio}\ \emph
  {et~al.}({\natexlab{b}})\citenamefont {Inguscio}, \citenamefont {Ketterle},\
  and\ \citenamefont {Salomon}}]{Varenna2006book}%
  \BibitemOpen
  \bibinfo {editor} {\bibfnamefont {M.}~\bibnamefont {Inguscio}}, \bibinfo
  {editor} {\bibfnamefont {W.}~\bibnamefont {Ketterle}},\ and\ \bibinfo
  {editor} {\bibfnamefont {C.}~\bibnamefont {Salomon}},\ eds.,\ \href@noop {}
  {\emph {\bibinfo {title} {{Ultra-cold Fermi Gases}}}}\ (\bibinfo  {publisher}
  {IOS Press, Amsterdam, 2008})\ \bibinfo {note} {{P}roceedings of the
  International School of Physics ``Enrico Fermi'', Course CLXIV, Varenna,
  20-30 June 2006}\BibitemShut {NoStop}%
\bibitem [{\citenamefont {Grimm}\ \emph {et~al.}()\citenamefont {Grimm},
  \citenamefont {Inguscio}, \citenamefont {Stringari},\ and\ \citenamefont
  {Lamporesi}}]{Varenna2022book}%
  \BibitemOpen
  \bibinfo {editor} {\bibfnamefont {R.}~\bibnamefont {Grimm}}, \bibinfo
  {editor} {\bibfnamefont {M.}~\bibnamefont {Inguscio}}, \bibinfo {editor}
  {\bibfnamefont {S.}~\bibnamefont {Stringari}},\ and\ \bibinfo {editor}
  {\bibfnamefont {G.}~\bibnamefont {Lamporesi}},\ eds.,\ \href@noop {} {\emph
  {\bibinfo {title} {{Quantum Mixtures with Ultra-cold Atoms}}}}\ (\bibinfo
  {publisher} {IOS Press, Amsterdam, 2025})\ \bibinfo {note} {{P}roceedings of
  the International School of Physics ``Enrico Fermi'', Course CCXI, Varenna,
  18-23 July 2022}\BibitemShut {NoStop}%
\bibitem [{\citenamefont {Weber}\ \emph {et~al.}(2003)\citenamefont {Weber},
  \citenamefont {Herbig}, \citenamefont {Mark}, \citenamefont {N\"agerl},\ and\
  \citenamefont {Grimm}}]{Weber2003bec}%
  \BibitemOpen
  \bibfield  {author} {\bibinfo {author} {\bibfnamefont {T.}~\bibnamefont
  {Weber}}, \bibinfo {author} {\bibfnamefont {J.}~\bibnamefont {Herbig}},
  \bibinfo {author} {\bibfnamefont {M.}~\bibnamefont {Mark}}, \bibinfo {author}
  {\bibfnamefont {H.-C.}\ \bibnamefont {N\"agerl}},\ and\ \bibinfo {author}
  {\bibfnamefont {R.}~\bibnamefont {Grimm}},\ }\bibfield  {title} {\bibinfo
  {title} {{Bose-Einstein} {C}ondensation of {C}esium},\ }\href
  {https://doi.org/10.1126/science.1079699} {\bibfield  {journal} {\bibinfo
  {journal} {Science}\ }\textbf {\bibinfo {volume} {299}},\ \bibinfo {pages}
  {232} (\bibinfo {year} {2003})}\BibitemShut {NoStop}%
\bibitem [{\citenamefont {Takasu}\ \emph {et~al.}(2003)\citenamefont {Takasu},
  \citenamefont {Maki}, \citenamefont {Komori}, \citenamefont {Takano},
  \citenamefont {Honda}, \citenamefont {Kumakura}, \citenamefont {Yabuzaki},\
  and\ \citenamefont {Takahashi}}]{Takasu2003ssb}%
  \BibitemOpen
  \bibfield  {author} {\bibinfo {author} {\bibfnamefont {Y.}~\bibnamefont
  {Takasu}}, \bibinfo {author} {\bibfnamefont {K.}~\bibnamefont {Maki}},
  \bibinfo {author} {\bibfnamefont {K.}~\bibnamefont {Komori}}, \bibinfo
  {author} {\bibfnamefont {T.}~\bibnamefont {Takano}}, \bibinfo {author}
  {\bibfnamefont {K.}~\bibnamefont {Honda}}, \bibinfo {author} {\bibfnamefont
  {M.}~\bibnamefont {Kumakura}}, \bibinfo {author} {\bibfnamefont
  {T.}~\bibnamefont {Yabuzaki}},\ and\ \bibinfo {author} {\bibfnamefont
  {Y.}~\bibnamefont {Takahashi}},\ }\bibfield  {title} {\bibinfo {title}
  {{Spin-singlet Bose-Einstein condensation of two-electron atoms}},\ }\href
  {https://doi.org/10.1103/PhysRevLett.91.040404} {\bibfield  {journal}
  {\bibinfo  {journal} {Phys. Rev. Lett.}\ }\textbf {\bibinfo {volume} {91}},\
  \bibinfo {pages} {040404} (\bibinfo {year} {2003})}\BibitemShut {NoStop}%
\bibitem [{\citenamefont {Griesmaier}\ \emph {et~al.}(2005)\citenamefont
  {Griesmaier}, \citenamefont {Werner}, \citenamefont {Hensler}, \citenamefont
  {Stuhler},\ and\ \citenamefont {Pfau}}]{Griesmaier2005bec}%
  \BibitemOpen
  \bibfield  {author} {\bibinfo {author} {\bibfnamefont {A.}~\bibnamefont
  {Griesmaier}}, \bibinfo {author} {\bibfnamefont {J.}~\bibnamefont {Werner}},
  \bibinfo {author} {\bibfnamefont {S.}~\bibnamefont {Hensler}}, \bibinfo
  {author} {\bibfnamefont {J.}~\bibnamefont {Stuhler}},\ and\ \bibinfo {author}
  {\bibfnamefont {T.}~\bibnamefont {Pfau}},\ }\bibfield  {title} {\bibinfo
  {title} {{Bose-Einstein condensation of chromium}},\ }\href
  {https://doi.org/10.1103/PhysRevLett.94.160401} {\bibfield  {journal}
  {\bibinfo  {journal} {Phys. Rev. Lett.}\ }\textbf {\bibinfo {volume} {94}},\
  \bibinfo {pages} {160401} (\bibinfo {year} {2005})}\BibitemShut {NoStop}%
\bibitem [{\citenamefont {Jones}\ \emph {et~al.}(2006)\citenamefont {Jones},
  \citenamefont {Tiesinga}, \citenamefont {Lett},\ and\ \citenamefont
  {Julienne}}]{Jones2006ups}%
  \BibitemOpen
  \bibfield  {author} {\bibinfo {author} {\bibfnamefont {K.~M.}\ \bibnamefont
  {Jones}}, \bibinfo {author} {\bibfnamefont {E.}~\bibnamefont {Tiesinga}},
  \bibinfo {author} {\bibfnamefont {P.~D.}\ \bibnamefont {Lett}},\ and\
  \bibinfo {author} {\bibfnamefont {P.~S.}\ \bibnamefont {Julienne}},\
  }\bibfield  {title} {\bibinfo {title} {Ultracold photoassociation
  spectroscopy: Long-range molecules and atomic scattering},\ }\href
  {https://doi.org/10.1103/RevModPhys.78.483} {\bibfield  {journal} {\bibinfo
  {journal} {Rev. Mod. Phys.}\ }\textbf {\bibinfo {volume} {78}},\ \bibinfo
  {eid} {483} (\bibinfo {year} {2006})}\BibitemShut {NoStop}%
\bibitem [{\citenamefont {Deiglmayr}\ \emph {et~al.}(2008)\citenamefont
  {Deiglmayr}, \citenamefont {Grochola}, \citenamefont {Repp}, \citenamefont
  {M\"{o}rtlbauer}, \citenamefont {Gl\"{u}ck}, \citenamefont {Lange},
  \citenamefont {Dulieu}, \citenamefont {Wester},\ and\ \citenamefont
  {Weidem\"{u}ller}}]{Deiglmayr2008fou}%
  \BibitemOpen
  \bibfield  {author} {\bibinfo {author} {\bibfnamefont {J.}~\bibnamefont
  {Deiglmayr}}, \bibinfo {author} {\bibfnamefont {A.}~\bibnamefont {Grochola}},
  \bibinfo {author} {\bibfnamefont {M.}~\bibnamefont {Repp}}, \bibinfo {author}
  {\bibfnamefont {K.}~\bibnamefont {M\"{o}rtlbauer}}, \bibinfo {author}
  {\bibfnamefont {C.}~\bibnamefont {Gl\"{u}ck}}, \bibinfo {author}
  {\bibfnamefont {J.}~\bibnamefont {Lange}}, \bibinfo {author} {\bibfnamefont
  {O.}~\bibnamefont {Dulieu}}, \bibinfo {author} {\bibfnamefont
  {R.}~\bibnamefont {Wester}},\ and\ \bibinfo {author} {\bibfnamefont
  {M.}~\bibnamefont {Weidem\"{u}ller}},\ }\bibfield  {title} {\bibinfo {title}
  {Formation of ultracold polar molecules in the rovibrational ground state},\
  }\href {https://doi.org/10.1103/PhysRevLett.101.133004} {\bibfield  {journal}
  {\bibinfo  {journal} {Phys. Rev. Lett.}\ }\textbf {\bibinfo {volume} {101}},\
  \bibinfo {eid} {133004} (\bibinfo {year} {2008})}\BibitemShut {NoStop}%
\bibitem [{\citenamefont {K\"ohler}\ \emph {et~al.}(2006)\citenamefont
  {K\"ohler}, \citenamefont {Goral},\ and\ \citenamefont
  {Julienne}}]{Koehler2006poc}%
  \BibitemOpen
  \bibfield  {author} {\bibinfo {author} {\bibfnamefont {T.}~\bibnamefont
  {K\"ohler}}, \bibinfo {author} {\bibfnamefont {K.}~\bibnamefont {Goral}},\
  and\ \bibinfo {author} {\bibfnamefont {P.~S.}\ \bibnamefont {Julienne}},\
  }\bibfield  {title} {\bibinfo {title} {Production of cold molecules via
  magnetically tunable {F}eshbach resonances},\ }\href
  {https://doi.org/10.1103/RevModPhys.78.1311} {\bibfield  {journal} {\bibinfo
  {journal} {Rev. Mod. Phys.}\ }\textbf {\bibinfo {volume} {78}},\ \bibinfo
  {pages} {1311} (\bibinfo {year} {2006})}\BibitemShut {NoStop}%
\bibitem [{\citenamefont {Chin}\ \emph {et~al.}(2010)\citenamefont {Chin},
  \citenamefont {Grimm}, \citenamefont {Julienne},\ and\ \citenamefont
  {Tiesinga}}]{Chin2010fri}%
  \BibitemOpen
  \bibfield  {author} {\bibinfo {author} {\bibfnamefont {C.}~\bibnamefont
  {Chin}}, \bibinfo {author} {\bibfnamefont {R.}~\bibnamefont {Grimm}},
  \bibinfo {author} {\bibfnamefont {P.~S.}\ \bibnamefont {Julienne}},\ and\
  \bibinfo {author} {\bibfnamefont {E.}~\bibnamefont {Tiesinga}},\ }\bibfield
  {title} {\bibinfo {title} {Feshbach resonances in ultracold gases},\ }\href
  {https://doi.org/doi.org/10.1103/RevModPhys.82.1225} {\bibfield  {journal}
  {\bibinfo  {journal} {Rev. Mod. Phys.}\ }\textbf {\bibinfo {volume} {82}},\
  \bibinfo {pages} {1225} (\bibinfo {year} {2010})}\BibitemShut {NoStop}%
\bibitem [{\citenamefont {Chin}\ \emph {et~al.}(2005)\citenamefont {Chin},
  \citenamefont {Kraemer}, \citenamefont {Mark}, \citenamefont {Herbig},
  \citenamefont {Waldburger}, \citenamefont {N\"agerl},\ and\ \citenamefont
  {Grimm}}]{Chin2005oof}%
  \BibitemOpen
  \bibfield  {author} {\bibinfo {author} {\bibfnamefont {C.}~\bibnamefont
  {Chin}}, \bibinfo {author} {\bibfnamefont {T.}~\bibnamefont {Kraemer}},
  \bibinfo {author} {\bibfnamefont {M.}~\bibnamefont {Mark}}, \bibinfo {author}
  {\bibfnamefont {J.}~\bibnamefont {Herbig}}, \bibinfo {author} {\bibfnamefont
  {P.}~\bibnamefont {Waldburger}}, \bibinfo {author} {\bibfnamefont {H.-C.}\
  \bibnamefont {N\"agerl}},\ and\ \bibinfo {author} {\bibfnamefont
  {R.}~\bibnamefont {Grimm}},\ }\bibfield  {title} {\bibinfo {title}
  {{Observation of Feshbach-like resonances in collisions between ultracold
  molecules}},\ }\href {https://doi.org/10.1103/PhysRevLett.94.123201}
  {\bibfield  {journal} {\bibinfo  {journal} {Phys. Rev. Lett.}\ }\textbf
  {\bibinfo {volume} {94}},\ \bibinfo {pages} {123201} (\bibinfo {year}
  {2005})}\BibitemShut {NoStop}%
\bibitem [{\citenamefont {Ferlaino}\ \emph {et~al.}(2008)\citenamefont
  {Ferlaino}, \citenamefont {Knoop}, \citenamefont {Mark}, \citenamefont
  {Berninger}, \citenamefont {Sch\"{o}bel}, \citenamefont {N\"{a}gerl},\ and\
  \citenamefont {Grimm}}]{Ferlaino2008cbt}%
  \BibitemOpen
  \bibfield  {author} {\bibinfo {author} {\bibfnamefont {F.}~\bibnamefont
  {Ferlaino}}, \bibinfo {author} {\bibfnamefont {S.}~\bibnamefont {Knoop}},
  \bibinfo {author} {\bibfnamefont {M.}~\bibnamefont {Mark}}, \bibinfo {author}
  {\bibfnamefont {M.}~\bibnamefont {Berninger}}, \bibinfo {author}
  {\bibfnamefont {H.}~\bibnamefont {Sch\"{o}bel}}, \bibinfo {author}
  {\bibfnamefont {H.-C.}\ \bibnamefont {N\"{a}gerl}},\ and\ \bibinfo {author}
  {\bibfnamefont {R.}~\bibnamefont {Grimm}},\ }\bibfield  {title} {\bibinfo
  {title} {{Collisions between tunable halo dimers: Exploring an elementary
  four-body process with identical bosons}},\ }\href
  {https://doi.org/10.1103/PhysRevLett.101.023201} {\bibfield  {journal}
  {\bibinfo  {journal} {Phys. Rev. Lett.}\ }\textbf {\bibinfo {volume} {101}},\
  \bibinfo {eid} {023201} (\bibinfo {year} {2008})}\BibitemShut {NoStop}%
\bibitem [{\citenamefont {Ni}\ \emph {et~al.}(2010)\citenamefont {Ni},
  \citenamefont {Ospelkaus}, \citenamefont {Wang}, \citenamefont {Quemener},
  \citenamefont {Neyenhuis}, \citenamefont {{de Miranda}}, \citenamefont
  {Bohn}, \citenamefont {Ye},\ and\ \citenamefont {Jin}}]{Ni2010dco}%
  \BibitemOpen
  \bibfield  {author} {\bibinfo {author} {\bibfnamefont {K.~K.}\ \bibnamefont
  {Ni}}, \bibinfo {author} {\bibfnamefont {S.}~\bibnamefont {Ospelkaus}},
  \bibinfo {author} {\bibfnamefont {D.}~\bibnamefont {Wang}}, \bibinfo {author}
  {\bibfnamefont {G.}~\bibnamefont {Quemener}}, \bibinfo {author}
  {\bibfnamefont {B.}~\bibnamefont {Neyenhuis}}, \bibinfo {author}
  {\bibfnamefont {M.~H.~G.}\ \bibnamefont {{de Miranda}}}, \bibinfo {author}
  {\bibfnamefont {J.~L.}\ \bibnamefont {Bohn}}, \bibinfo {author}
  {\bibfnamefont {J.}~\bibnamefont {Ye}},\ and\ \bibinfo {author}
  {\bibfnamefont {D.~S.}\ \bibnamefont {Jin}},\ }\bibfield  {title} {\bibinfo
  {title} {Dipolar collisions of polar molecules in the quantum regime},\
  }\href {https://doi.org/10.1038/nature08953} {\bibfield  {journal} {\bibinfo
  {journal} {Nature (London)}\ }\textbf {\bibinfo {volume} {464}},\ \bibinfo
  {pages} {1324} (\bibinfo {year} {2010})}\BibitemShut {NoStop}%
\bibitem [{\citenamefont {Mark}\ \emph {et~al.}(2007)\citenamefont {Mark},
  \citenamefont {Ferlaino}, \citenamefont {Knoop}, \citenamefont {Danzl},
  \citenamefont {Kraemer}, \citenamefont {Chin}, \citenamefont {N\"{a}gerl},\
  and\ \citenamefont {Grimm}}]{Mark2007sou}%
  \BibitemOpen
  \bibfield  {author} {\bibinfo {author} {\bibfnamefont {M.}~\bibnamefont
  {Mark}}, \bibinfo {author} {\bibfnamefont {F.}~\bibnamefont {Ferlaino}},
  \bibinfo {author} {\bibfnamefont {S.}~\bibnamefont {Knoop}}, \bibinfo
  {author} {\bibfnamefont {J.~G.}\ \bibnamefont {Danzl}}, \bibinfo {author}
  {\bibfnamefont {T.}~\bibnamefont {Kraemer}}, \bibinfo {author} {\bibfnamefont
  {C.}~\bibnamefont {Chin}}, \bibinfo {author} {\bibfnamefont {H.-C.}\
  \bibnamefont {N\"{a}gerl}},\ and\ \bibinfo {author} {\bibfnamefont
  {R.}~\bibnamefont {Grimm}},\ }\bibfield  {title} {\bibinfo {title}
  {Spectroscopy of ultracold trapped cesium {F}eshbach molecules},\ }\href
  {https://doi.org/10.1103/PhysRevA.76.042514} {\bibfield  {journal} {\bibinfo
  {journal} {Phys. Rev. A}\ }\textbf {\bibinfo {volume} {76}},\ \bibinfo
  {pages} {042514} (\bibinfo {year} {2007})}\BibitemShut {NoStop}%
\bibitem [{\citenamefont {Winkler}\ \emph {et~al.}(2007)\citenamefont
  {Winkler}, \citenamefont {Lang}, \citenamefont {Thalhammer}, \citenamefont
  {{van der Straten}}, \citenamefont {Grimm},\ and\ \citenamefont {{Hecker
  Denschlag}}}]{Winkler2007cot}%
  \BibitemOpen
  \bibfield  {author} {\bibinfo {author} {\bibfnamefont {K.}~\bibnamefont
  {Winkler}}, \bibinfo {author} {\bibfnamefont {F.}~\bibnamefont {Lang}},
  \bibinfo {author} {\bibfnamefont {G.}~\bibnamefont {Thalhammer}}, \bibinfo
  {author} {\bibfnamefont {P.}~\bibnamefont {{van der Straten}}}, \bibinfo
  {author} {\bibfnamefont {R.}~\bibnamefont {Grimm}},\ and\ \bibinfo {author}
  {\bibfnamefont {J.}~\bibnamefont {{Hecker Denschlag}}},\ }\bibfield  {title}
  {\bibinfo {title} {{Coherent optical transfer of Feshbach molecules to a
  lower vibrational state}},\ }\href
  {https://doi.org/10.1103/PhysRevLett.98.043201} {\bibfield  {journal}
  {\bibinfo  {journal} {Phys. Rev. Lett.}\ }\textbf {\bibinfo {volume} {98}},\
  \bibinfo {eid} {043201} (\bibinfo {year} {2007})}\BibitemShut {NoStop}%
\bibitem [{\citenamefont {Lang}\ \emph
  {et~al.}(2008{\natexlab{a}})\citenamefont {Lang}, \citenamefont {{van der
  Straten}}, \citenamefont {Brandst\"atter}, \citenamefont {Thalhammer},
  \citenamefont {Winkler}, \citenamefont {Julienne}, \citenamefont {Grimm},\
  and\ \citenamefont {Hecker~Denschlag}}]{Lang2008ctm}%
  \BibitemOpen
  \bibfield  {author} {\bibinfo {author} {\bibfnamefont {F.}~\bibnamefont
  {Lang}}, \bibinfo {author} {\bibfnamefont {P.}~\bibnamefont {{van der
  Straten}}}, \bibinfo {author} {\bibfnamefont {B.}~\bibnamefont
  {Brandst\"atter}}, \bibinfo {author} {\bibfnamefont {G.}~\bibnamefont
  {Thalhammer}}, \bibinfo {author} {\bibfnamefont {K.}~\bibnamefont {Winkler}},
  \bibinfo {author} {\bibfnamefont {P.~S.}\ \bibnamefont {Julienne}}, \bibinfo
  {author} {\bibfnamefont {R.}~\bibnamefont {Grimm}},\ and\ \bibinfo {author}
  {\bibfnamefont {J.}~\bibnamefont {Hecker~Denschlag}},\ }\bibfield  {title}
  {\bibinfo {title} {Cruising through molecular bound-state manifolds with
  radiofrequency},\ }\href {https://doi.org/https://doi.org/10.1038/nphys838}
  {\bibfield  {journal} {\bibinfo  {journal} {Nat. Phys.}\ }\textbf {\bibinfo
  {volume} {4}},\ \bibinfo {pages} {223} (\bibinfo {year}
  {2008}{\natexlab{a}})}\BibitemShut {NoStop}%
\bibitem [{\citenamefont {Ni}\ \emph {et~al.}(2008)\citenamefont {Ni},
  \citenamefont {Ospelkaus}, \citenamefont {de~Miranda}, \citenamefont {Pe'er},
  \citenamefont {Neyenhuis}, \citenamefont {Zirbel}, \citenamefont
  {Kotochigova}, \citenamefont {Julienne}, \citenamefont {Jin},\ and\
  \citenamefont {Ye}}]{Ni2008ahp}%
  \BibitemOpen
  \bibfield  {author} {\bibinfo {author} {\bibfnamefont {K.-K.}\ \bibnamefont
  {Ni}}, \bibinfo {author} {\bibfnamefont {S.}~\bibnamefont {Ospelkaus}},
  \bibinfo {author} {\bibfnamefont {M.~H.~G.}\ \bibnamefont {de~Miranda}},
  \bibinfo {author} {\bibfnamefont {A.}~\bibnamefont {Pe'er}}, \bibinfo
  {author} {\bibfnamefont {B.}~\bibnamefont {Neyenhuis}}, \bibinfo {author}
  {\bibfnamefont {J.~J.}\ \bibnamefont {Zirbel}}, \bibinfo {author}
  {\bibfnamefont {S.}~\bibnamefont {Kotochigova}}, \bibinfo {author}
  {\bibfnamefont {P.~S.}\ \bibnamefont {Julienne}}, \bibinfo {author}
  {\bibfnamefont {D.~S.}\ \bibnamefont {Jin}},\ and\ \bibinfo {author}
  {\bibfnamefont {J.}~\bibnamefont {Ye}},\ }\bibfield  {title} {\bibinfo
  {title} {{A High Phase-Space-Density Gas of Polar Molecules}},\ }\href
  {https://doi.org/10.1126/science.1163861} {\bibfield  {journal} {\bibinfo
  {journal} {Science}\ }\textbf {\bibinfo {volume} {322}},\ \bibinfo {pages}
  {231} (\bibinfo {year} {2008})}\BibitemShut {NoStop}%
\bibitem [{\citenamefont {Lang}\ \emph
  {et~al.}(2008{\natexlab{b}})\citenamefont {Lang}, \citenamefont {Winkler},
  \citenamefont {Strauss}, \citenamefont {Grimm},\ and\ \citenamefont {{Hecker
  Denschlag}}}]{Lang2008utm}%
  \BibitemOpen
  \bibfield  {author} {\bibinfo {author} {\bibfnamefont {F.}~\bibnamefont
  {Lang}}, \bibinfo {author} {\bibfnamefont {K.}~\bibnamefont {Winkler}},
  \bibinfo {author} {\bibfnamefont {C.}~\bibnamefont {Strauss}}, \bibinfo
  {author} {\bibfnamefont {R.}~\bibnamefont {Grimm}},\ and\ \bibinfo {author}
  {\bibfnamefont {J.}~\bibnamefont {{Hecker Denschlag}}},\ }\bibfield  {title}
  {\bibinfo {title} {Ultracold triplet molecules in the rovibrational ground
  state},\ }\href {https://doi.org/10.1103/PhysRevLett.101.133005} {\bibfield
  {journal} {\bibinfo  {journal} {Phys. Rev. Lett.}\ }\textbf {\bibinfo
  {volume} {101}},\ \bibinfo {eid} {133005} (\bibinfo {year}
  {2008}{\natexlab{b}})}\BibitemShut {NoStop}%
\bibitem [{\citenamefont {Danzl}\ \emph {et~al.}(2010)\citenamefont {Danzl},
  \citenamefont {Mark}, \citenamefont {Haller}, \citenamefont {Gustavsson},
  \citenamefont {Hart}, \citenamefont {Aldegunde}, \citenamefont {Hutson},\
  and\ \citenamefont {N\"{a}gerl}}]{Danzl2010auh}%
  \BibitemOpen
  \bibfield  {author} {\bibinfo {author} {\bibfnamefont {J.~G.}\ \bibnamefont
  {Danzl}}, \bibinfo {author} {\bibfnamefont {M.~J.}\ \bibnamefont {Mark}},
  \bibinfo {author} {\bibfnamefont {E.}~\bibnamefont {Haller}}, \bibinfo
  {author} {\bibfnamefont {M.}~\bibnamefont {Gustavsson}}, \bibinfo {author}
  {\bibfnamefont {R.}~\bibnamefont {Hart}}, \bibinfo {author} {\bibfnamefont
  {J.}~\bibnamefont {Aldegunde}}, \bibinfo {author} {\bibfnamefont {J.~M.}\
  \bibnamefont {Hutson}},\ and\ \bibinfo {author} {\bibfnamefont {H.-C.}\
  \bibnamefont {N\"{a}gerl}},\ }\bibfield  {title} {\bibinfo {title} {An
  ultracold high-density sample of rovibronic ground-state molecules in an
  optical lattice},\ }\href {https://doi.org/10.1038/nphys1533} {\bibfield
  {journal} {\bibinfo  {journal} {Nat. Phys.}\ }\textbf {\bibinfo {volume}
  {6}},\ \bibinfo {pages} {265} (\bibinfo {year} {2010})}\BibitemShut {NoStop}%
\bibitem [{\citenamefont {Jochim}\ \emph
  {et~al.}(2003{\natexlab{a}})\citenamefont {Jochim}, \citenamefont
  {Bartenstein}, \citenamefont {Altmeyer}, \citenamefont {Hendl}, \citenamefont
  {Riedl}, \citenamefont {Chin}, \citenamefont {{Hecker Denschlag}},\ and\
  \citenamefont {Grimm}}]{Jochim2003bec}%
  \BibitemOpen
  \bibfield  {author} {\bibinfo {author} {\bibfnamefont {S.}~\bibnamefont
  {Jochim}}, \bibinfo {author} {\bibfnamefont {M.}~\bibnamefont {Bartenstein}},
  \bibinfo {author} {\bibfnamefont {A.}~\bibnamefont {Altmeyer}}, \bibinfo
  {author} {\bibfnamefont {G.}~\bibnamefont {Hendl}}, \bibinfo {author}
  {\bibfnamefont {S.}~\bibnamefont {Riedl}}, \bibinfo {author} {\bibfnamefont
  {C.}~\bibnamefont {Chin}}, \bibinfo {author} {\bibfnamefont {J.}~\bibnamefont
  {{Hecker Denschlag}}},\ and\ \bibinfo {author} {\bibfnamefont
  {R.}~\bibnamefont {Grimm}},\ }\bibfield  {title} {\bibinfo {title}
  {{Bose-Einstein Condensation of Molecules}},\ }\href
  {https://doi.org/10.1126/science.1093280} {\bibfield  {journal} {\bibinfo
  {journal} {Science}\ }\textbf {\bibinfo {volume} {302}},\ \bibinfo {pages}
  {2101} (\bibinfo {year} {2003}{\natexlab{a}})}\BibitemShut {NoStop}%
\bibitem [{\citenamefont {Greiner}\ \emph {et~al.}(2003)\citenamefont
  {Greiner}, \citenamefont {Regal},\ and\ \citenamefont
  {Jin}}]{Greiner2003eoa}%
  \BibitemOpen
  \bibfield  {author} {\bibinfo {author} {\bibfnamefont {M.}~\bibnamefont
  {Greiner}}, \bibinfo {author} {\bibfnamefont {C.~A.}\ \bibnamefont {Regal}},\
  and\ \bibinfo {author} {\bibfnamefont {D.~S.}\ \bibnamefont {Jin}},\
  }\bibfield  {title} {\bibinfo {title} {{Emergence of a molecular
  Bose-Einstein condensate from a Fermi gas}},\ }\href
  {https://doi.org/doi.org/10.1038/nature02199} {\bibfield  {journal} {\bibinfo
   {journal} {Nature (London)}\ }\textbf {\bibinfo {volume} {426}},\ \bibinfo
  {pages} {537} (\bibinfo {year} {2003})}\BibitemShut {NoStop}%
\bibitem [{\citenamefont {Zwierlein}\ \emph {et~al.}(2003)\citenamefont
  {Zwierlein}, \citenamefont {Stan}, \citenamefont {Schunck}, \citenamefont
  {Raupach}, \citenamefont {Gupta}, \citenamefont {Hadzibabic},\ and\
  \citenamefont {Ketterle}}]{Zwierlein2003oob}%
  \BibitemOpen
  \bibfield  {author} {\bibinfo {author} {\bibfnamefont {M.~W.}\ \bibnamefont
  {Zwierlein}}, \bibinfo {author} {\bibfnamefont {C.~A.}\ \bibnamefont {Stan}},
  \bibinfo {author} {\bibfnamefont {C.~H.}\ \bibnamefont {Schunck}}, \bibinfo
  {author} {\bibfnamefont {S.~M.~F.}\ \bibnamefont {Raupach}}, \bibinfo
  {author} {\bibfnamefont {S.}~\bibnamefont {Gupta}}, \bibinfo {author}
  {\bibfnamefont {Z.}~\bibnamefont {Hadzibabic}},\ and\ \bibinfo {author}
  {\bibfnamefont {W.}~\bibnamefont {Ketterle}},\ }\bibfield  {title} {\bibinfo
  {title} {{Observation of Bose-Einstein Condensation of Molecules}},\ }\href
  {https://doi.org/10.1103/PhysRevLett.91.250401} {\bibfield  {journal}
  {\bibinfo  {journal} {Phys. Rev. Lett.}\ }\textbf {\bibinfo {volume} {91}},\
  \bibinfo {pages} {250401} (\bibinfo {year} {2003})}\BibitemShut {NoStop}%
\bibitem [{\citenamefont {Bartenstein}\ \emph {et~al.}(2004)\citenamefont
  {Bartenstein}, \citenamefont {Altmeyer}, \citenamefont {Riedl}, \citenamefont
  {Jochim}, \citenamefont {Chin}, \citenamefont {{Hecker Denschlag}},\ and\
  \citenamefont {Grimm}}]{Bartenstein2004cfa}%
  \BibitemOpen
  \bibfield  {author} {\bibinfo {author} {\bibfnamefont {M.}~\bibnamefont
  {Bartenstein}}, \bibinfo {author} {\bibfnamefont {A.}~\bibnamefont
  {Altmeyer}}, \bibinfo {author} {\bibfnamefont {S.}~\bibnamefont {Riedl}},
  \bibinfo {author} {\bibfnamefont {S.}~\bibnamefont {Jochim}}, \bibinfo
  {author} {\bibfnamefont {C.}~\bibnamefont {Chin}}, \bibinfo {author}
  {\bibfnamefont {J.}~\bibnamefont {{Hecker Denschlag}}},\ and\ \bibinfo
  {author} {\bibfnamefont {R.}~\bibnamefont {Grimm}},\ }\bibfield  {title}
  {\bibinfo {title} {{Crossover from a Molecular Bose-Einstein Condensate to a
  Degenerate Fermi Gas}},\ }\href
  {https://doi.org/10.1103/PhysRevLett.92.120401} {\bibfield  {journal}
  {\bibinfo  {journal} {Phys. Rev. Lett.}\ }\textbf {\bibinfo {volume} {92}},\
  \bibinfo {pages} {120401} (\bibinfo {year} {2004})}\BibitemShut {NoStop}%
\bibitem [{\citenamefont {Regal}\ \emph
  {et~al.}(2004{\natexlab{a}})\citenamefont {Regal}, \citenamefont {Greiner},\
  and\ \citenamefont {Jin}}]{Regal2004oor}%
  \BibitemOpen
  \bibfield  {author} {\bibinfo {author} {\bibfnamefont {C.~A.}\ \bibnamefont
  {Regal}}, \bibinfo {author} {\bibfnamefont {M.}~\bibnamefont {Greiner}},\
  and\ \bibinfo {author} {\bibfnamefont {D.~S.}\ \bibnamefont {Jin}},\
  }\bibfield  {title} {\bibinfo {title} {{Observation of Resonance Condensation
  of Fermionic Atom Pairs}},\ }\href
  {https://doi.org/10.1103/PhysRevLett.92.040403} {\bibfield  {journal}
  {\bibinfo  {journal} {Phys. Rev. Lett.}\ }\textbf {\bibinfo {volume} {92}},\
  \bibinfo {pages} {040403} (\bibinfo {year} {2004}{\natexlab{a}})}\BibitemShut
  {NoStop}%
\bibitem [{\citenamefont {Zwierlein}\ \emph {et~al.}(2004)\citenamefont
  {Zwierlein}, \citenamefont {Stan}, \citenamefont {Schunck}, \citenamefont
  {Raupach}, \citenamefont {Kerman},\ and\ \citenamefont
  {Ketterle}}]{Zwierlein2004cop}%
  \BibitemOpen
  \bibfield  {author} {\bibinfo {author} {\bibfnamefont {M.~W.}\ \bibnamefont
  {Zwierlein}}, \bibinfo {author} {\bibfnamefont {C.~A.}\ \bibnamefont {Stan}},
  \bibinfo {author} {\bibfnamefont {C.~H.}\ \bibnamefont {Schunck}}, \bibinfo
  {author} {\bibfnamefont {S.~M.~F.}\ \bibnamefont {Raupach}}, \bibinfo
  {author} {\bibfnamefont {A.~J.}\ \bibnamefont {Kerman}},\ and\ \bibinfo
  {author} {\bibfnamefont {W.}~\bibnamefont {Ketterle}},\ }\bibfield  {title}
  {\bibinfo {title} {{Condensation of Pairs of Fermionic Atoms near a Feshbach
  Resonance}},\ }\href {https://doi.org/10.1103/PhysRevLett.92.120403}
  {\bibfield  {journal} {\bibinfo  {journal} {Phys. Rev. Lett.}\ }\textbf
  {\bibinfo {volume} {92}},\ \bibinfo {pages} {120403} (\bibinfo {year}
  {2004})}\BibitemShut {NoStop}%
\bibitem [{\citenamefont {Zwerger}(2012)}]{Zwerger2012tbb}%
  \BibitemOpen
  \bibinfo {editor} {\bibfnamefont {W.}~\bibnamefont {Zwerger}},\ ed.,\
  \href@noop {} {\emph {\bibinfo {title} {The BCS-BEC Crossover and the Unitary
  Fermi Gas}}}\ (\bibinfo  {publisher} {Springer, Berlin Heidelberg},\ \bibinfo
  {year} {2012})\BibitemShut {NoStop}%
\bibitem [{\citenamefont {Strinati}\ \emph {et~al.}(2018)\citenamefont
  {Strinati}, \citenamefont {Pieri}, \citenamefont {R{\"o}pke}, \citenamefont
  {Schuck},\ and\ \citenamefont {Urban}}]{Strinati2018tbb}%
  \BibitemOpen
  \bibfield  {author} {\bibinfo {author} {\bibfnamefont {G.~C.}\ \bibnamefont
  {Strinati}}, \bibinfo {author} {\bibfnamefont {P.}~\bibnamefont {Pieri}},
  \bibinfo {author} {\bibfnamefont {G.}~\bibnamefont {R{\"o}pke}}, \bibinfo
  {author} {\bibfnamefont {P.}~\bibnamefont {Schuck}},\ and\ \bibinfo {author}
  {\bibfnamefont {M.}~\bibnamefont {Urban}},\ }\bibfield  {title} {\bibinfo
  {title} {The {BCS-BEC} crossover: From ultra-cold {F}ermi gases to nuclear
  systems},\ }\href {https://doi.org/10.1016/j.physrep.2018.02.004} {\bibfield
  {journal} {\bibinfo  {journal} {Phys. Rep.}\ }\textbf {\bibinfo {volume}
  {738}},\ \bibinfo {pages} {1} (\bibinfo {year} {2018})}\BibitemShut {NoStop}%
\bibitem [{\citenamefont {Zhang}\ \emph {et~al.}(2021)\citenamefont {Zhang},
  \citenamefont {Chen}, \citenamefont {Yao},\ and\ \citenamefont
  {Chin}}]{Zhang2021tfa}%
  \BibitemOpen
  \bibfield  {author} {\bibinfo {author} {\bibfnamefont {Z.}~\bibnamefont
  {Zhang}}, \bibinfo {author} {\bibfnamefont {L.}~\bibnamefont {Chen}},
  \bibinfo {author} {\bibfnamefont {K.-X.}\ \bibnamefont {Yao}},\ and\ \bibinfo
  {author} {\bibfnamefont {C.}~\bibnamefont {Chin}},\ }\bibfield  {title}
  {\bibinfo {title} {Transition from an atomic to a molecular
  {B}ose–{E}instein condensate},\ }\href
  {https://doi.org/10.1038/s41586-021-03443-0} {\bibfield  {journal} {\bibinfo
  {journal} {Nature (London)}\ }\textbf {\bibinfo {volume} {592}},\ \bibinfo
  {pages} {708} (\bibinfo {year} {2021})}\BibitemShut {NoStop}%
\bibitem [{\citenamefont {Bigagli}\ \emph {et~al.}(2024)\citenamefont
  {Bigagli}, \citenamefont {Yuan}, \citenamefont {Zhang}, \citenamefont
  {Bulatovic}, \citenamefont {Karman}, \citenamefont {Stevenson},\ and\
  \citenamefont {Will}}]{Bigagli2024oob}%
  \BibitemOpen
  \bibfield  {author} {\bibinfo {author} {\bibfnamefont {N.}~\bibnamefont
  {Bigagli}}, \bibinfo {author} {\bibfnamefont {W.}~\bibnamefont {Yuan}},
  \bibinfo {author} {\bibfnamefont {S.}~\bibnamefont {Zhang}}, \bibinfo
  {author} {\bibfnamefont {B.}~\bibnamefont {Bulatovic}}, \bibinfo {author}
  {\bibfnamefont {T.}~\bibnamefont {Karman}}, \bibinfo {author} {\bibfnamefont
  {I.}~\bibnamefont {Stevenson}},\ and\ \bibinfo {author} {\bibfnamefont
  {S.}~\bibnamefont {Will}},\ }\bibfield  {title} {\bibinfo {title}
  {{Observation of Bose–Einstein condensation of dipolar molecules}},\ }\href
  {https://doi.org/10.1038/s41586-024-07492-z} {\bibfield  {journal} {\bibinfo
  {journal} {Nature (London)}\ }\textbf {\bibinfo {volume} {631}},\ \bibinfo
  {pages} {289} (\bibinfo {year} {2024})}\BibitemShut {NoStop}%
\bibitem [{\citenamefont {Shi}\ \emph {et~al.}(2025)\citenamefont {Shi},
  \citenamefont {Huang}, \citenamefont {Deng}, \citenamefont {Jin},
  \citenamefont {Yi}, \citenamefont {Shi},\ and\ \citenamefont
  {Wang}}]{Shi2025bec}%
  \BibitemOpen
  \bibfield  {author} {\bibinfo {author} {\bibfnamefont {Z.}~\bibnamefont
  {Shi}}, \bibinfo {author} {\bibfnamefont {Z.}~\bibnamefont {Huang}}, \bibinfo
  {author} {\bibfnamefont {F.}~\bibnamefont {Deng}}, \bibinfo {author}
  {\bibfnamefont {W.-J.}\ \bibnamefont {Jin}}, \bibinfo {author} {\bibfnamefont
  {S.}~\bibnamefont {Yi}}, \bibinfo {author} {\bibfnamefont {T.}~\bibnamefont
  {Shi}},\ and\ \bibinfo {author} {\bibfnamefont {D.}~\bibnamefont {Wang}},\
  }\href {https://arxiv.org/abs/2508.20518} {\bibinfo {title} {Bose-einstein
  condensate of ultracold sodium-rubidium molecules with tunable dipolar
  interactions}} (\bibinfo {year} {2025}),\ \Eprint
  {https://arxiv.org/abs/2508.20518} {arXiv:2508.20518 [cond-mat.quant-gas]}
  \BibitemShut {NoStop}%
\bibitem [{\citenamefont {Zhang}\ \emph {et~al.}(2020)\citenamefont {Zhang},
  \citenamefont {Yu}, \citenamefont {Cairncross}, \citenamefont {Wang},
  \citenamefont {Picard}, \citenamefont {Hood}, \citenamefont {Lin},
  \citenamefont {Hutson},\ and\ \citenamefont {Ni}}]{Zhang2020fas}%
  \BibitemOpen
  \bibfield  {author} {\bibinfo {author} {\bibfnamefont {J.~T.}\ \bibnamefont
  {Zhang}}, \bibinfo {author} {\bibfnamefont {Y.}~\bibnamefont {Yu}}, \bibinfo
  {author} {\bibfnamefont {W.~B.}\ \bibnamefont {Cairncross}}, \bibinfo
  {author} {\bibfnamefont {K.}~\bibnamefont {Wang}}, \bibinfo {author}
  {\bibfnamefont {L.~R.~B.}\ \bibnamefont {Picard}}, \bibinfo {author}
  {\bibfnamefont {J.~D.}\ \bibnamefont {Hood}}, \bibinfo {author}
  {\bibfnamefont {Y.-W.}\ \bibnamefont {Lin}}, \bibinfo {author} {\bibfnamefont
  {J.~M.}\ \bibnamefont {Hutson}},\ and\ \bibinfo {author} {\bibfnamefont
  {K.-K.}\ \bibnamefont {Ni}},\ }\bibfield  {title} {\bibinfo {title} {{Forming
  a Single Molecule by Magnetoassociation in an Optical Tweezer}},\ }\href
  {https://doi.org/10.1103/PhysRevLett.124.253401} {\bibfield  {journal}
  {\bibinfo  {journal} {Phys. Rev. Lett.}\ }\textbf {\bibinfo {volume} {124}},\
  \bibinfo {pages} {253401} (\bibinfo {year} {2020})}\BibitemShut {NoStop}%
\bibitem [{\citenamefont {Cairncross}\ \emph {et~al.}(2021)\citenamefont
  {Cairncross}, \citenamefont {Zhang}, \citenamefont {Picard}, \citenamefont
  {Yu}, \citenamefont {Wang},\ and\ \citenamefont {Ni}}]{Cairncross2021aoa}%
  \BibitemOpen
  \bibfield  {author} {\bibinfo {author} {\bibfnamefont {W.~B.}\ \bibnamefont
  {Cairncross}}, \bibinfo {author} {\bibfnamefont {J.~T.}\ \bibnamefont
  {Zhang}}, \bibinfo {author} {\bibfnamefont {L.~R.~B.}\ \bibnamefont
  {Picard}}, \bibinfo {author} {\bibfnamefont {Y.}~\bibnamefont {Yu}}, \bibinfo
  {author} {\bibfnamefont {K.}~\bibnamefont {Wang}},\ and\ \bibinfo {author}
  {\bibfnamefont {K.-K.}\ \bibnamefont {Ni}},\ }\bibfield  {title} {\bibinfo
  {title} {Assembly of a rovibrational ground state molecule in an optical
  tweezer},\ }\href {https://doi.org/10.1103/PhysRevLett.126.123402} {\bibfield
   {journal} {\bibinfo  {journal} {Phys. Rev. Lett.}\ }\textbf {\bibinfo
  {volume} {126}},\ \bibinfo {pages} {123402} (\bibinfo {year}
  {2021})}\BibitemShut {NoStop}%
\bibitem [{\citenamefont {Ruttley}\ \emph {et~al.}(2023)\citenamefont
  {Ruttley}, \citenamefont {Guttridge}, \citenamefont {Spence}, \citenamefont
  {Bird}, \citenamefont {Le~Sueur}, \citenamefont {Hutson},\ and\ \citenamefont
  {Cornish}}]{Ruttley2023fou}%
  \BibitemOpen
  \bibfield  {author} {\bibinfo {author} {\bibfnamefont {D.~K.}\ \bibnamefont
  {Ruttley}}, \bibinfo {author} {\bibfnamefont {A.}~\bibnamefont {Guttridge}},
  \bibinfo {author} {\bibfnamefont {S.}~\bibnamefont {Spence}}, \bibinfo
  {author} {\bibfnamefont {R.~C.}\ \bibnamefont {Bird}}, \bibinfo {author}
  {\bibfnamefont {C.~R.}\ \bibnamefont {Le~Sueur}}, \bibinfo {author}
  {\bibfnamefont {J.~M.}\ \bibnamefont {Hutson}},\ and\ \bibinfo {author}
  {\bibfnamefont {S.~L.}\ \bibnamefont {Cornish}},\ }\bibfield  {title}
  {\bibinfo {title} {Formation of ultracold molecules by merging optical
  tweezers},\ }\href {https://doi.org/10.1103/PhysRevLett.130.223401}
  {\bibfield  {journal} {\bibinfo  {journal} {Phys. Rev. Lett.}\ }\textbf
  {\bibinfo {volume} {130}},\ \bibinfo {pages} {223401} (\bibinfo {year}
  {2023})}\BibitemShut {NoStop}%
\bibitem [{\citenamefont {Anderegg}\ \emph {et~al.}(2019)\citenamefont
  {Anderegg}, \citenamefont {Cheuk}, \citenamefont {Bao}, \citenamefont
  {Burchesky}, \citenamefont {Ketterle}, \citenamefont {Ni},\ and\
  \citenamefont {Doyle}}]{Anderegg2019aot}%
  \BibitemOpen
  \bibfield  {author} {\bibinfo {author} {\bibfnamefont {L.}~\bibnamefont
  {Anderegg}}, \bibinfo {author} {\bibfnamefont {L.~W.}\ \bibnamefont {Cheuk}},
  \bibinfo {author} {\bibfnamefont {Y.}~\bibnamefont {Bao}}, \bibinfo {author}
  {\bibfnamefont {S.}~\bibnamefont {Burchesky}}, \bibinfo {author}
  {\bibfnamefont {W.}~\bibnamefont {Ketterle}}, \bibinfo {author}
  {\bibfnamefont {K.-K.}\ \bibnamefont {Ni}},\ and\ \bibinfo {author}
  {\bibfnamefont {J.~M.}\ \bibnamefont {Doyle}},\ }\bibfield  {title} {\bibinfo
  {title} {An optical tweezer array of ultracold molecules},\ }\href
  {https://doi.org/10.1126/science.aax1265} {\bibfield  {journal} {\bibinfo
  {journal} {Science}\ }\textbf {\bibinfo {volume} {365}},\ \bibinfo {pages}
  {1156} (\bibinfo {year} {2019})}\BibitemShut {NoStop}%
\bibitem [{\citenamefont {Vilas}\ \emph {et~al.}(2024)\citenamefont {Vilas},
  \citenamefont {Robichaud}, \citenamefont {Hallas}, \citenamefont {Li},
  \citenamefont {Anderegg},\ and\ \citenamefont {Doyle}}]{Vilas2024aot}%
  \BibitemOpen
  \bibfield  {author} {\bibinfo {author} {\bibfnamefont {N.~B.}\ \bibnamefont
  {Vilas}}, \bibinfo {author} {\bibfnamefont {P.}~\bibnamefont {Robichaud}},
  \bibinfo {author} {\bibfnamefont {C.}~\bibnamefont {Hallas}}, \bibinfo
  {author} {\bibfnamefont {G.~K.}\ \bibnamefont {Li}}, \bibinfo {author}
  {\bibfnamefont {L.}~\bibnamefont {Anderegg}},\ and\ \bibinfo {author}
  {\bibfnamefont {J.~M.}\ \bibnamefont {Doyle}},\ }\bibfield  {title} {\bibinfo
  {title} {An optical tweezer array of ultracold polyatomic molecules},\ }\href
  {https://doi.org/10.1038/s41586-024-07199-1} {\bibfield  {journal} {\bibinfo
  {journal} {Nature (London)}\ }\textbf {\bibinfo {volume} {628}},\ \bibinfo
  {pages} {282} (\bibinfo {year} {2024})}\BibitemShut {NoStop}%
\bibitem [{\citenamefont {Soave}\ \emph {et~al.}(2023)\citenamefont {Soave},
  \citenamefont {Canali}, \citenamefont {Ye}, \citenamefont {Kreyer},
  \citenamefont {Kirilov},\ and\ \citenamefont {Grimm}}]{Soave2023otf}%
  \BibitemOpen
  \bibfield  {author} {\bibinfo {author} {\bibfnamefont {E.}~\bibnamefont
  {Soave}}, \bibinfo {author} {\bibfnamefont {A.}~\bibnamefont {Canali}},
  \bibinfo {author} {\bibfnamefont {Z.-X.}\ \bibnamefont {Ye}}, \bibinfo
  {author} {\bibfnamefont {M.}~\bibnamefont {Kreyer}}, \bibinfo {author}
  {\bibfnamefont {E.}~\bibnamefont {Kirilov}},\ and\ \bibinfo {author}
  {\bibfnamefont {R.}~\bibnamefont {Grimm}},\ }\bibfield  {title} {\bibinfo
  {title} {{Optically trapped Feshbach molecules of fermionic
  $^{161}\mathrm{Dy}$ and $^{40}\mathrm{K}$}},\ }\href
  {https://doi.org/10.1103/PhysRevResearch.5.033117} {\bibfield  {journal}
  {\bibinfo  {journal} {Phys. Rev. Res.}\ }\textbf {\bibinfo {volume} {5}},\
  \bibinfo {pages} {033117} (\bibinfo {year} {2023})}\BibitemShut {NoStop}%
\bibitem [{\citenamefont {Ciamei}\ \emph {et~al.}(2022)\citenamefont {Ciamei},
  \citenamefont {Finelli}, \citenamefont {Cosco}, \citenamefont {Inguscio},
  \citenamefont {Trenkwalder},\ and\ \citenamefont {Zaccanti}}]{Ciamei2022ddf}%
  \BibitemOpen
  \bibfield  {author} {\bibinfo {author} {\bibfnamefont {A.}~\bibnamefont
  {Ciamei}}, \bibinfo {author} {\bibfnamefont {S.}~\bibnamefont {Finelli}},
  \bibinfo {author} {\bibfnamefont {A.}~\bibnamefont {Cosco}}, \bibinfo
  {author} {\bibfnamefont {M.}~\bibnamefont {Inguscio}}, \bibinfo {author}
  {\bibfnamefont {A.}~\bibnamefont {Trenkwalder}},\ and\ \bibinfo {author}
  {\bibfnamefont {M.}~\bibnamefont {Zaccanti}},\ }\bibfield  {title} {\bibinfo
  {title} {{Double-degenerate Fermi mixtures of $^{6}\mathrm{Li}$ and
  $^{53}\mathrm{Cr}$ atoms}},\ }\href
  {https://doi.org/10.1103/PhysRevA.106.053318} {\bibfield  {journal} {\bibinfo
   {journal} {Phys. Rev. A}\ }\textbf {\bibinfo {volume} {106}},\ \bibinfo
  {pages} {053318} (\bibinfo {year} {2022})}\BibitemShut {NoStop}%
\bibitem [{\citenamefont {Ravensbergen}\ \emph {et~al.}(2020)\citenamefont
  {Ravensbergen}, \citenamefont {Soave}, \citenamefont {Corre}, \citenamefont
  {Kreyer}, \citenamefont {Huang}, \citenamefont {Kirilov},\ and\ \citenamefont
  {Grimm}}]{Ravensbergen2020rif}%
  \BibitemOpen
  \bibfield  {author} {\bibinfo {author} {\bibfnamefont {C.}~\bibnamefont
  {Ravensbergen}}, \bibinfo {author} {\bibfnamefont {E.}~\bibnamefont {Soave}},
  \bibinfo {author} {\bibfnamefont {V.}~\bibnamefont {Corre}}, \bibinfo
  {author} {\bibfnamefont {M.}~\bibnamefont {Kreyer}}, \bibinfo {author}
  {\bibfnamefont {B.}~\bibnamefont {Huang}}, \bibinfo {author} {\bibfnamefont
  {E.}~\bibnamefont {Kirilov}},\ and\ \bibinfo {author} {\bibfnamefont
  {R.}~\bibnamefont {Grimm}},\ }\bibfield  {title} {\bibinfo {title}
  {{Resonantly Interacting Fermi-Fermi Mixture of $^{161}\mathrm{Dy}$ and
  $^{40}\mathrm{K}$}},\ }\href {https://doi.org/10.1103/PhysRevLett.124.203402}
  {\bibfield  {journal} {\bibinfo  {journal} {Phys. Rev. Lett.}\ }\textbf
  {\bibinfo {volume} {124}},\ \bibinfo {pages} {203402} (\bibinfo {year}
  {2020})}\BibitemShut {NoStop}%
\bibitem [{\citenamefont {Ye}\ \emph {et~al.}(2025)\citenamefont {Ye},
  \citenamefont {Canali}, \citenamefont {Wong}, \citenamefont {Kreyer},
  \citenamefont {Kirilov},\ and\ \citenamefont {Grimm}}]{Ye2025dms}%
  \BibitemOpen
  \bibfield  {author} {\bibinfo {author} {\bibfnamefont {Z.-X.}\ \bibnamefont
  {Ye}}, \bibinfo {author} {\bibfnamefont {A.}~\bibnamefont {Canali}}, \bibinfo
  {author} {\bibfnamefont {C.-K.}\ \bibnamefont {Wong}}, \bibinfo {author}
  {\bibfnamefont {M.}~\bibnamefont {Kreyer}}, \bibinfo {author} {\bibfnamefont
  {E.}~\bibnamefont {Kirilov}},\ and\ \bibinfo {author} {\bibfnamefont
  {R.}~\bibnamefont {Grimm}},\ }\bibfield  {title} {\bibinfo {title}
  {Dipole-mode spectrum and hydrodynamic crossover in a resonantly interacting
  two-species fermion mixture},\ }\href
  {https://doi.org/10.1103/PhysRevResearch.7.023259} {\bibfield  {journal}
  {\bibinfo  {journal} {Phys. Rev. Res.}\ }\textbf {\bibinfo {volume} {7}},\
  \bibinfo {pages} {023259} (\bibinfo {year} {2025})}\BibitemShut {NoStop}%
\bibitem [{\citenamefont {Gubbels}\ \emph {et~al.}(2009)\citenamefont
  {Gubbels}, \citenamefont {Baarsma},\ and\ \citenamefont
  {Stoof}}]{Gubbels2009lpi}%
  \BibitemOpen
  \bibfield  {author} {\bibinfo {author} {\bibfnamefont {K.~B.}\ \bibnamefont
  {Gubbels}}, \bibinfo {author} {\bibfnamefont {J.~E.}\ \bibnamefont
  {Baarsma}},\ and\ \bibinfo {author} {\bibfnamefont {H.~T.~C.}\ \bibnamefont
  {Stoof}},\ }\bibfield  {title} {\bibinfo {title} {Lifshitz {P}oint in the
  {P}hase {D}iagram of {R}esonantly {I}nteracting
  $^{6}\mathrm{Li}\mathrm{\text{-}}^{40}\mathrm{K}$ {M}ixtures},\ }\href
  {https://doi.org/10.1103/PhysRevLett.103.195301} {\bibfield  {journal}
  {\bibinfo  {journal} {Phys. Rev. Lett.}\ }\textbf {\bibinfo {volume} {103}},\
  \bibinfo {pages} {195301} (\bibinfo {year} {2009})}\BibitemShut {NoStop}%
\bibitem [{\citenamefont {Gubbels}\ and\ \citenamefont
  {Stoof}(2013)}]{Gubbels2013ifg}%
  \BibitemOpen
  \bibfield  {author} {\bibinfo {author} {\bibfnamefont {K.~B.}\ \bibnamefont
  {Gubbels}}\ and\ \bibinfo {author} {\bibfnamefont {H.~T.~C.}\ \bibnamefont
  {Stoof}},\ }\bibfield  {title} {\bibinfo {title} {Imbalanced {F}ermi gases at
  unitarity},\ }\href {https://doi.org/10.1016/j.physrep.2012.11.004}
  {\bibfield  {journal} {\bibinfo  {journal} {Phys. Rep.}\ }\textbf {\bibinfo
  {volume} {525}},\ \bibinfo {pages} {255 } (\bibinfo {year}
  {2013})}\BibitemShut {NoStop}%
\bibitem [{\citenamefont {Wang}\ \emph {et~al.}(2017)\citenamefont {Wang},
  \citenamefont {Che}, \citenamefont {Zhang},\ and\ \citenamefont
  {Chen}}]{Wang2017eeo}%
  \BibitemOpen
  \bibfield  {author} {\bibinfo {author} {\bibfnamefont {J.}~\bibnamefont
  {Wang}}, \bibinfo {author} {\bibfnamefont {Y.}~\bibnamefont {Che}}, \bibinfo
  {author} {\bibfnamefont {L.}~\bibnamefont {Zhang}},\ and\ \bibinfo {author}
  {\bibfnamefont {Q.}~\bibnamefont {Chen}},\ }\bibfield  {title} {\bibinfo
  {title} {Enhancement effect of mass imbalance on
  {Fulde-Ferrell-Larkin-Ovchinnikov} type of pairing in {Fermi-Fermi} mixtures
  of ultracold quantum gases},\ }\href {https://doi.org/10.1038/srep39783}
  {\bibfield  {journal} {\bibinfo  {journal} {Sci. Rep.}\ }\textbf {\bibinfo
  {volume} {7}},\ \bibinfo {pages} {39783} (\bibinfo {year}
  {2017})}\BibitemShut {NoStop}%
\bibitem [{\citenamefont {Pini}\ \emph {et~al.}(2021)\citenamefont {Pini},
  \citenamefont {Pieri}, \citenamefont {Grimm},\ and\ \citenamefont
  {Strinati}}]{Pini2021bmf}%
  \BibitemOpen
  \bibfield  {author} {\bibinfo {author} {\bibfnamefont {M.}~\bibnamefont
  {Pini}}, \bibinfo {author} {\bibfnamefont {P.}~\bibnamefont {Pieri}},
  \bibinfo {author} {\bibfnamefont {R.}~\bibnamefont {Grimm}},\ and\ \bibinfo
  {author} {\bibfnamefont {G.~C.}\ \bibnamefont {Strinati}},\ }\bibfield
  {title} {\bibinfo {title} {{Beyond-mean-field description of a trapped
  unitary Fermi gas with mass and population imbalance}},\ }\href
  {https://doi.org/10.1103/PhysRevA.103.023314} {\bibfield  {journal} {\bibinfo
   {journal} {Phys. Rev. A}\ }\textbf {\bibinfo {volume} {103}},\ \bibinfo
  {pages} {023314} (\bibinfo {year} {2021})}\BibitemShut {NoStop}%
\bibitem [{\citenamefont {Fulde}\ and\ \citenamefont
  {Ferrell}(1964)}]{Fulde1964sia}%
  \BibitemOpen
  \bibfield  {author} {\bibinfo {author} {\bibfnamefont {P.}~\bibnamefont
  {Fulde}}\ and\ \bibinfo {author} {\bibfnamefont {R.~A.}\ \bibnamefont
  {Ferrell}},\ }\bibfield  {title} {\bibinfo {title} {{Superconductivity in a
  Strong Spin-Exchange Field}},\ }\href
  {https://doi.org/10.1103/PhysRev.135.A550} {\bibfield  {journal} {\bibinfo
  {journal} {Phys. Rev.}\ }\textbf {\bibinfo {volume} {135}},\ \bibinfo {pages}
  {A550} (\bibinfo {year} {1964})}\BibitemShut {NoStop}%
\bibitem [{\citenamefont {Larkin}\ and\ \citenamefont
  {Ovchinnikov}(1964)}]{Larkin1964nss}%
  \BibitemOpen
  \bibfield  {author} {\bibinfo {author} {\bibfnamefont {A.~I.}\ \bibnamefont
  {Larkin}}\ and\ \bibinfo {author} {\bibfnamefont {Y.~N.}\ \bibnamefont
  {Ovchinnikov}},\ }\bibfield  {title} {\bibinfo {title} {Neodnorodnoye
  sostoyanie sverkhprovodnikov},\ }\href@noop {} {\bibfield  {journal}
  {\bibinfo  {journal} {Zh. Eksp. Teor. Fiz.}\ }\textbf {\bibinfo {volume}
  {47}},\ \bibinfo {pages} {1136} (\bibinfo {year} {1964})},\ \bibinfo {note}
  {[Sov. Phys. JETP 20, 762 (1965)].}\BibitemShut {Stop}%
\bibitem [{\citenamefont {Radzihovsky}\ and\ \citenamefont
  {Sheehy}(2010)}]{Radzihovsky2010ifr}%
  \BibitemOpen
  \bibfield  {author} {\bibinfo {author} {\bibfnamefont {L.}~\bibnamefont
  {Radzihovsky}}\ and\ \bibinfo {author} {\bibfnamefont {D.~E.}\ \bibnamefont
  {Sheehy}},\ }\bibfield  {title} {\bibinfo {title} {{Imbalanced
  Feshbach-resonant Fermi gases}},\ }\href
  {https://doi.org/10.1088/0034-4885/73/7/076501} {\bibfield  {journal}
  {\bibinfo  {journal} {Rep. Prog. Phys.}\ }\textbf {\bibinfo {volume} {73}},\
  \bibinfo {pages} {076501} (\bibinfo {year} {2010})}\BibitemShut {NoStop}%
\bibitem [{\citenamefont {Naidon}\ and\ \citenamefont
  {Endo}(2017)}]{Naidon2017epa}%
  \BibitemOpen
  \bibfield  {author} {\bibinfo {author} {\bibfnamefont {P.}~\bibnamefont
  {Naidon}}\ and\ \bibinfo {author} {\bibfnamefont {S.}~\bibnamefont {Endo}},\
  }\bibfield  {title} {\bibinfo {title} {Efimov physics: a review},\ }\href
  {https://doi.org/10.1088/1361-6633/aa50e8} {\bibfield  {journal} {\bibinfo
  {journal} {Rep. Prog. Phys.}\ }\textbf {\bibinfo {volume} {80}},\ \bibinfo
  {pages} {056001} (\bibinfo {year} {2017})}\BibitemShut {NoStop}%
\bibitem [{\citenamefont {Kartavtsev}\ and\ \citenamefont
  {Malykh}(2007)}]{Kartavtsev2007let}%
  \BibitemOpen
  \bibfield  {author} {\bibinfo {author} {\bibfnamefont {O.~I.}\ \bibnamefont
  {Kartavtsev}}\ and\ \bibinfo {author} {\bibfnamefont {A.~V.}\ \bibnamefont
  {Malykh}},\ }\bibfield  {title} {\bibinfo {title} {Low-energy three-body
  dynamics in binary quantum gases},\ }\href
  {https://doi.org/10.1088/0953-4075/40/7/011} {\bibfield  {journal} {\bibinfo
  {journal} {J. Phys. B: At. Mol. Opt. Phys.}\ }\textbf {\bibinfo {volume}
  {40}},\ \bibinfo {pages} {1429} (\bibinfo {year} {2007})}\BibitemShut
  {NoStop}%
\bibitem [{\citenamefont {Zaccanti}(2025)}]{Zaccanti2022mif}%
  \BibitemOpen
  \bibfield  {author} {\bibinfo {author} {\bibfnamefont {M.}~\bibnamefont
  {Zaccanti}},\ }\bibfield  {title} {\bibinfo {title} {{Mass-imbalanced Fermi
  mixtures with resonant interactions}},\ }\href@noop {} {\bibfield  {journal}
  {\bibinfo  {journal} {in Ref.~\cite{Varenna2022book}}\ } (\bibinfo {year}
  {2025})}\BibitemShut {NoStop}%
\bibitem [{\citenamefont {Petrov}\ \emph {et~al.}(2004)\citenamefont {Petrov},
  \citenamefont {Salomon},\ and\ \citenamefont {Shlyapnikov}}]{Petrov2004wbd}%
  \BibitemOpen
  \bibfield  {author} {\bibinfo {author} {\bibfnamefont {D.~S.}\ \bibnamefont
  {Petrov}}, \bibinfo {author} {\bibfnamefont {C.}~\bibnamefont {Salomon}},\
  and\ \bibinfo {author} {\bibfnamefont {G.~V.}\ \bibnamefont {Shlyapnikov}},\
  }\bibfield  {title} {\bibinfo {title} {{Weakly Bound Dimers of Fermionic
  Atoms}},\ }\href {https://doi.org/10.1103/PhysRevLett.93.090404} {\bibfield
  {journal} {\bibinfo  {journal} {Phys. Rev. Lett.}\ }\textbf {\bibinfo
  {volume} {93}},\ \bibinfo {pages} {090404} (\bibinfo {year}
  {2004})}\BibitemShut {NoStop}%
\bibitem [{\citenamefont {Petrov}\ \emph {et~al.}(2005)\citenamefont {Petrov},
  \citenamefont {Salomon},\ and\ \citenamefont {Shlyapnikov}}]{Petrov2005dmi}%
  \BibitemOpen
  \bibfield  {author} {\bibinfo {author} {\bibfnamefont {D.~S.}\ \bibnamefont
  {Petrov}}, \bibinfo {author} {\bibfnamefont {C.}~\bibnamefont {Salomon}},\
  and\ \bibinfo {author} {\bibfnamefont {G.~V.}\ \bibnamefont {Shlyapnikov}},\
  }\bibfield  {title} {\bibinfo {title} {Diatomic molecules in ultracold
  {F}ermi gases -- novel composite bosons},\ }\href
  {https://doi.org/10.1088/0953-4075/38/9/014} {\bibfield  {journal} {\bibinfo
  {journal} {J. Phys. B}\ }\textbf {\bibinfo {volume} {38}},\ \bibinfo {pages}
  {S645} (\bibinfo {year} {2005})}\BibitemShut {NoStop}%
\bibitem [{\citenamefont {Jag}\ \emph {et~al.}(2016)\citenamefont {Jag},
  \citenamefont {Cetina}, \citenamefont {Lous}, \citenamefont {Grimm},
  \citenamefont {Levinsen},\ and\ \citenamefont {Petrov}}]{Jag2016lof}%
  \BibitemOpen
  \bibfield  {author} {\bibinfo {author} {\bibfnamefont {M.}~\bibnamefont
  {Jag}}, \bibinfo {author} {\bibfnamefont {M.}~\bibnamefont {Cetina}},
  \bibinfo {author} {\bibfnamefont {R.~S.}\ \bibnamefont {Lous}}, \bibinfo
  {author} {\bibfnamefont {R.}~\bibnamefont {Grimm}}, \bibinfo {author}
  {\bibfnamefont {J.}~\bibnamefont {Levinsen}},\ and\ \bibinfo {author}
  {\bibfnamefont {D.~S.}\ \bibnamefont {Petrov}},\ }\bibfield  {title}
  {\bibinfo {title} {Lifetime of {F}eshbach dimers in a {F}ermi-{F}ermi mixture
  of $^{6}\text{Li}$ and $^{40}\text{K}$},\ }\href
  {https://doi.org/10.1103/PhysRevA.94.062706} {\bibfield  {journal} {\bibinfo
  {journal} {Phys. Rev. A}\ }\textbf {\bibinfo {volume} {94}},\ \bibinfo
  {pages} {062706} (\bibinfo {year} {2016})}\BibitemShut {NoStop}%
\bibitem [{\citenamefont {Chotia}\ \emph {et~al.}(2012)\citenamefont {Chotia},
  \citenamefont {Neyenhuis}, \citenamefont {Moses}, \citenamefont {Yan},
  \citenamefont {Covey}, \citenamefont {Foss-Feig}, \citenamefont {Rey},
  \citenamefont {Jin},\ and\ \citenamefont {Ye}}]{Chotia2012lld}%
  \BibitemOpen
  \bibfield  {author} {\bibinfo {author} {\bibfnamefont {A.}~\bibnamefont
  {Chotia}}, \bibinfo {author} {\bibfnamefont {B.}~\bibnamefont {Neyenhuis}},
  \bibinfo {author} {\bibfnamefont {S.~A.}\ \bibnamefont {Moses}}, \bibinfo
  {author} {\bibfnamefont {B.}~\bibnamefont {Yan}}, \bibinfo {author}
  {\bibfnamefont {J.~P.}\ \bibnamefont {Covey}}, \bibinfo {author}
  {\bibfnamefont {M.}~\bibnamefont {Foss-Feig}}, \bibinfo {author}
  {\bibfnamefont {A.~M.}\ \bibnamefont {Rey}}, \bibinfo {author} {\bibfnamefont
  {D.~S.}\ \bibnamefont {Jin}},\ and\ \bibinfo {author} {\bibfnamefont
  {J.}~\bibnamefont {Ye}},\ }\bibfield  {title} {\bibinfo {title} {{Long-Lived
  Dipolar Molecules and Feshbach Molecules in a 3D Optical Lattice}},\ }\href
  {https://doi.org/10.1103/PhysRevLett.108.080405} {\bibfield  {journal}
  {\bibinfo  {journal} {Phys. Rev. Lett.}\ }\textbf {\bibinfo {volume} {108}},\
  \bibinfo {pages} {080405} (\bibinfo {year} {2012})}\BibitemShut {NoStop}%
\bibitem [{\citenamefont {Spence}(2023)}]{Spence2023ASR}%
  \BibitemOpen
  \bibfield  {author} {\bibinfo {author} {\bibfnamefont {S.~J.}\ \bibnamefont
  {Spence}},\ }\emph {\bibinfo {title} {Assembling Single RbCs Molecules with
  Optical Tweezers}},\ \href {http://etheses.dur.ac.uk/14814/} {Ph.D. thesis},\
  \bibinfo  {school} {Durham University}, \bibinfo {address} {Durham
  University} (\bibinfo {year} {2023})\BibitemShut {NoStop}%
\bibitem [{\citenamefont {Finelli}\ \emph {et~al.}(2024)\citenamefont
  {Finelli}, \citenamefont {Ciamei}, \citenamefont {Restivo}, \citenamefont
  {Schemmer}, \citenamefont {Cosco}, \citenamefont {Inguscio}, \citenamefont
  {Trenkwalder}, \citenamefont {Zaremba-Kopczyk}, \citenamefont {Gronowski},
  \citenamefont {Tomza},\ and\ \citenamefont {Zaccanti}}]{Finelli2024ula}%
  \BibitemOpen
  \bibfield  {author} {\bibinfo {author} {\bibfnamefont {S.}~\bibnamefont
  {Finelli}}, \bibinfo {author} {\bibfnamefont {A.}~\bibnamefont {Ciamei}},
  \bibinfo {author} {\bibfnamefont {B.}~\bibnamefont {Restivo}}, \bibinfo
  {author} {\bibfnamefont {M.}~\bibnamefont {Schemmer}}, \bibinfo {author}
  {\bibfnamefont {A.}~\bibnamefont {Cosco}}, \bibinfo {author} {\bibfnamefont
  {M.}~\bibnamefont {Inguscio}}, \bibinfo {author} {\bibfnamefont
  {A.}~\bibnamefont {Trenkwalder}}, \bibinfo {author} {\bibfnamefont
  {K.}~\bibnamefont {Zaremba-Kopczyk}}, \bibinfo {author} {\bibfnamefont
  {M.}~\bibnamefont {Gronowski}}, \bibinfo {author} {\bibfnamefont
  {M.}~\bibnamefont {Tomza}},\ and\ \bibinfo {author} {\bibfnamefont
  {M.}~\bibnamefont {Zaccanti}},\ }\bibfield  {title} {\bibinfo {title}
  {{Ultracold $\mathrm{Li}\mathrm{Cr}$: A New Pathway to Quantum Gases of
  Paramagnetic Polar Molecules}},\ }\href
  {https://doi.org/10.1103/PRXQuantum.5.020358} {\bibfield  {journal} {\bibinfo
   {journal} {PRX Quantum}\ }\textbf {\bibinfo {volume} {5}},\ \bibinfo {pages}
  {020358} (\bibinfo {year} {2024})}\BibitemShut {NoStop}%
\bibitem [{\citenamefont {Ye}\ \emph {et~al.}(2022)\citenamefont {Ye},
  \citenamefont {Canali}, \citenamefont {Soave}, \citenamefont {Kreyer},
  \citenamefont {Yudkin}, \citenamefont {Ravensbergen}, \citenamefont
  {Kirilov},\ and\ \citenamefont {Grimm}}]{Ye2022ool}%
  \BibitemOpen
  \bibfield  {author} {\bibinfo {author} {\bibfnamefont {Z.-X.}\ \bibnamefont
  {Ye}}, \bibinfo {author} {\bibfnamefont {A.}~\bibnamefont {Canali}}, \bibinfo
  {author} {\bibfnamefont {E.}~\bibnamefont {Soave}}, \bibinfo {author}
  {\bibfnamefont {M.}~\bibnamefont {Kreyer}}, \bibinfo {author} {\bibfnamefont
  {Y.}~\bibnamefont {Yudkin}}, \bibinfo {author} {\bibfnamefont
  {C.}~\bibnamefont {Ravensbergen}}, \bibinfo {author} {\bibfnamefont
  {E.}~\bibnamefont {Kirilov}},\ and\ \bibinfo {author} {\bibfnamefont
  {R.}~\bibnamefont {Grimm}},\ }\bibfield  {title} {\bibinfo {title}
  {{Observation of low-field Feshbach resonances between $^{161}\mathrm{Dy}$
  and $^{40}\mathrm{K}$}},\ }\href
  {https://doi.org/10.1103/PhysRevA.106.043314} {\bibfield  {journal} {\bibinfo
   {journal} {Phys. Rev. A}\ }\textbf {\bibinfo {volume} {106}},\ \bibinfo
  {pages} {043314} (\bibinfo {year} {2022})}\BibitemShut {NoStop}%
\bibitem [{\citenamefont {Petrov}(2004)}]{Petrov2004tbp}%
  \BibitemOpen
  \bibfield  {author} {\bibinfo {author} {\bibfnamefont {D.~S.}\ \bibnamefont
  {Petrov}},\ }\bibfield  {title} {\bibinfo {title} {{Three-Boson Problem near
  a Narrow {F}eshbach Resonance}},\ }\href
  {https://doi.org/10.1103/PhysRevLett.93.143201} {\bibfield  {journal}
  {\bibinfo  {journal} {Phys. Rev. Lett.}\ }\textbf {\bibinfo {volume} {93}},\
  \bibinfo {pages} {143201} (\bibinfo {year} {2004})}\BibitemShut {NoStop}%
\bibitem [{\citenamefont {Braaten}\ and\ \citenamefont
  {Hammer}(2006)}]{Braaten2006uif}%
  \BibitemOpen
  \bibfield  {author} {\bibinfo {author} {\bibfnamefont {E.}~\bibnamefont
  {Braaten}}\ and\ \bibinfo {author} {\bibfnamefont {H.-W.}\ \bibnamefont
  {Hammer}},\ }\bibfield  {title} {\bibinfo {title} {Universality in few-body
  systems with large scattering length},\ }\href
  {https://doi.org/10.1016/j.physrep.2006.03.001} {\bibfield  {journal}
  {\bibinfo  {journal} {Phys. Rep.}\ }\textbf {\bibinfo {volume} {428}},\
  \bibinfo {pages} {259} (\bibinfo {year} {2006})}\BibitemShut {NoStop}%
\bibitem [{\citenamefont {Ravensbergen}\ \emph
  {et~al.}(2018{\natexlab{b}})\citenamefont {Ravensbergen}, \citenamefont
  {Corre}, \citenamefont {Soave}, \citenamefont {Kreyer}, \citenamefont
  {Kiri\-lov},\ and\ \citenamefont {Grimm}}]{Ravensbergen2018poa}%
  \BibitemOpen
  \bibfield  {author} {\bibinfo {author} {\bibfnamefont {C.}~\bibnamefont
  {Ravensbergen}}, \bibinfo {author} {\bibfnamefont {V.}~\bibnamefont {Corre}},
  \bibinfo {author} {\bibfnamefont {E.}~\bibnamefont {Soave}}, \bibinfo
  {author} {\bibfnamefont {M.}~\bibnamefont {Kreyer}}, \bibinfo {author}
  {\bibfnamefont {E.}~\bibnamefont {Kiri\-lov}},\ and\ \bibinfo {author}
  {\bibfnamefont {R.}~\bibnamefont {Grimm}},\ }\bibfield  {title} {\bibinfo
  {title} {Production of a degenerate {F}ermi-{F}ermi mixture of dysprosium and
  potassium atoms},\ }\href {https://doi.org/10.1103/PhysRevA.98.063624}
  {\bibfield  {journal} {\bibinfo  {journal} {Phys. Rev. A}\ }\textbf {\bibinfo
  {volume} {98}},\ \bibinfo {pages} {063624} (\bibinfo {year}
  {2018}{\natexlab{b}})},\ \bibinfo {note} {{\it ibid.} {\bf 101}, 059903(E)
  (2020).}\BibitemShut {Stop}%
\bibitem [{\citenamefont {Thompson}\ \emph {et~al.}(2005)\citenamefont
  {Thompson}, \citenamefont {Hodby},\ and\ \citenamefont
  {Wieman}}]{Thompson2005sdo}%
  \BibitemOpen
  \bibfield  {author} {\bibinfo {author} {\bibfnamefont {S.~T.}\ \bibnamefont
  {Thompson}}, \bibinfo {author} {\bibfnamefont {E.}~\bibnamefont {Hodby}},\
  and\ \bibinfo {author} {\bibfnamefont {C.~E.}\ \bibnamefont {Wieman}},\
  }\bibfield  {title} {\bibinfo {title} {{Spontaneous Dissociation of $^{85}$Rb
  Feshbach Molecules}},\ }\href {https://doi.org/10.1103/PhysRevLett.94.020401}
  {\bibfield  {journal} {\bibinfo  {journal} {Phys. Rev. Lett.}\ }\textbf
  {\bibinfo {volume} {94}},\ \bibinfo {pages} {020401} (\bibinfo {year}
  {2005})}\BibitemShut {NoStop}%
\bibitem [{Note1()}]{Note1}%
  \BibitemOpen
  \bibinfo {note} {The potassium imaging provides a better quality signal
  compared to the Dy ones, due to the higher absorption cross
  section.}\BibitemShut {Stop}%
\bibitem [{\citenamefont {Zaheer}\ \emph {et~al.}()\citenamefont {Zaheer},
  \citenamefont {Bouras}, \citenamefont {Giergiel}, \citenamefont {Simoni},
  \citenamefont {Grimm}, \citenamefont {Sidorov},\ and\ \citenamefont
  {Hannaford}}]{Melbourne}%
  \BibitemOpen
  \bibfield  {author} {\bibinfo {author} {\bibfnamefont {A.}~\bibnamefont
  {Zaheer}}, \bibinfo {author} {\bibfnamefont {M.}~\bibnamefont {Bouras}},
  \bibinfo {author} {\bibfnamefont {K.}~\bibnamefont {Giergiel}}, \bibinfo
  {author} {\bibfnamefont {A.}~\bibnamefont {Simoni}}, \bibinfo {author}
  {\bibfnamefont {R.}~\bibnamefont {Grimm}}, \bibinfo {author} {\bibfnamefont
  {A.}~\bibnamefont {Sidorov}},\ and\ \bibinfo {author} {\bibfnamefont
  {P.}~\bibnamefont {Hannaford}},\ }\href@noop {} {\bibinfo {title} {Large
  anomalous shifts of potassium-39 {F}eshbach resonances, manuscript in
  preparation}}\BibitemShut {NoStop}%
\bibitem [{\citenamefont {Kasahara}\ \emph {et~al.}(1999)\citenamefont
  {Kasahara}, \citenamefont {Fujiwara}, \citenamefont {Okada}, \citenamefont
  {Katô},\ and\ \citenamefont {Baba}}]{Kasahara1999dfo}%
  \BibitemOpen
  \bibfield  {author} {\bibinfo {author} {\bibfnamefont {S.}~\bibnamefont
  {Kasahara}}, \bibinfo {author} {\bibfnamefont {C.}~\bibnamefont {Fujiwara}},
  \bibinfo {author} {\bibfnamefont {N.}~\bibnamefont {Okada}}, \bibinfo
  {author} {\bibfnamefont {H.}~\bibnamefont {Katô}},\ and\ \bibinfo {author}
  {\bibfnamefont {M.}~\bibnamefont {Baba}},\ }\bibfield  {title} {\bibinfo
  {title} {Doppler-free optical–optical double resonance polarization
  spectroscopy of the $^{39}\mathrm{K}^{85}\mathrm{Rb}$ $1^1\mathrm{\Pi}$ and
  2$^1\mathrm{\Pi}$ states},\ }\href {https://doi.org/10.1063/1.480256}
  {\bibfield  {journal} {\bibinfo  {journal} {J. Chem. Phys.}\ }\textbf
  {\bibinfo {volume} {111}},\ \bibinfo {pages} {8857} (\bibinfo {year}
  {1999})}\BibitemShut {NoStop}%
\bibitem [{\citenamefont {Guo}\ \emph {et~al.}(2016)\citenamefont {Guo},
  \citenamefont {Zhu}, \citenamefont {Lu}, \citenamefont {Ye}, \citenamefont
  {Wang}, \citenamefont {Vexiau}, \citenamefont {Bouloufa-Maafa}, \citenamefont
  {Qu\'em\'ener}, \citenamefont {Dulieu},\ and\ \citenamefont
  {Wang}}]{Guo2016coa}%
  \BibitemOpen
  \bibfield  {author} {\bibinfo {author} {\bibfnamefont {M.}~\bibnamefont
  {Guo}}, \bibinfo {author} {\bibfnamefont {B.}~\bibnamefont {Zhu}}, \bibinfo
  {author} {\bibfnamefont {B.}~\bibnamefont {Lu}}, \bibinfo {author}
  {\bibfnamefont {X.}~\bibnamefont {Ye}}, \bibinfo {author} {\bibfnamefont
  {F.}~\bibnamefont {Wang}}, \bibinfo {author} {\bibfnamefont {R.}~\bibnamefont
  {Vexiau}}, \bibinfo {author} {\bibfnamefont {N.}~\bibnamefont
  {Bouloufa-Maafa}}, \bibinfo {author} {\bibfnamefont {G.}~\bibnamefont
  {Qu\'em\'ener}}, \bibinfo {author} {\bibfnamefont {O.}~\bibnamefont
  {Dulieu}},\ and\ \bibinfo {author} {\bibfnamefont {D.}~\bibnamefont {Wang}},\
  }\bibfield  {title} {\bibinfo {title} {Creation of an ultracold gas of
  ground-state dipolar $^{23}\mathrm{Na}^{87}\mathrm{Rb}$ molecules},\ }\href
  {https://doi.org/10.1103/PhysRevLett.116.205303} {\bibfield  {journal}
  {\bibinfo  {journal} {Phys. Rev. Lett.}\ }\textbf {\bibinfo {volume} {116}},\
  \bibinfo {pages} {205303} (\bibinfo {year} {2016})}\BibitemShut {NoStop}%
\bibitem [{\citenamefont {Guo}\ \emph {et~al.}(2017)\citenamefont {Guo},
  \citenamefont {Vexiau}, \citenamefont {Zhu}, \citenamefont {Lu},
  \citenamefont {Bouloufa-Maafa}, \citenamefont {Dulieu},\ and\ \citenamefont
  {Wang}}]{Guo2017hrm}%
  \BibitemOpen
  \bibfield  {author} {\bibinfo {author} {\bibfnamefont {M.}~\bibnamefont
  {Guo}}, \bibinfo {author} {\bibfnamefont {R.}~\bibnamefont {Vexiau}},
  \bibinfo {author} {\bibfnamefont {B.}~\bibnamefont {Zhu}}, \bibinfo {author}
  {\bibfnamefont {B.}~\bibnamefont {Lu}}, \bibinfo {author} {\bibfnamefont
  {N.}~\bibnamefont {Bouloufa-Maafa}}, \bibinfo {author} {\bibfnamefont
  {O.}~\bibnamefont {Dulieu}},\ and\ \bibinfo {author} {\bibfnamefont
  {D.}~\bibnamefont {Wang}},\ }\bibfield  {title} {\bibinfo {title}
  {High-resolution molecular spectroscopy for producing ultracold
  absolute-ground-state $^{23}\mathrm{Na}^{87}\mathrm{Rb}$ molecules},\ }\href
  {https://doi.org/10.1103/PhysRevA.96.052505} {\bibfield  {journal} {\bibinfo
  {journal} {Phys. Rev. A}\ }\textbf {\bibinfo {volume} {96}},\ \bibinfo
  {pages} {052505} (\bibinfo {year} {2017})}\BibitemShut {NoStop}%
\bibitem [{\citenamefont {Dzuba}\ \emph {et~al.}(2011)\citenamefont {Dzuba},
  \citenamefont {Flambaum},\ and\ \citenamefont {Lev}}]{Dzuba2011dpa}%
  \BibitemOpen
  \bibfield  {author} {\bibinfo {author} {\bibfnamefont {V.~A.}\ \bibnamefont
  {Dzuba}}, \bibinfo {author} {\bibfnamefont {V.~V.}\ \bibnamefont
  {Flambaum}},\ and\ \bibinfo {author} {\bibfnamefont {B.~L.}\ \bibnamefont
  {Lev}},\ }\bibfield  {title} {\bibinfo {title} {Dynamic polarizabilities and
  magic wavelengths for dysprosium},\ }\href
  {https://doi.org/10.1103/PhysRevA.83.032502} {\bibfield  {journal} {\bibinfo
  {journal} {Phys. Rev. A}\ }\textbf {\bibinfo {volume} {83}},\ \bibinfo
  {pages} {032502} (\bibinfo {year} {2011})}\BibitemShut {NoStop}%
\bibitem [{\citenamefont {Li}\ \emph {et~al.}(2017)\citenamefont {Li},
  \citenamefont {Wyart}, \citenamefont {Dulieu}, \citenamefont
  {Nascimb{\`e}ne},\ and\ \citenamefont {Lepers}}]{Li2017oto}%
  \BibitemOpen
  \bibfield  {author} {\bibinfo {author} {\bibfnamefont {H.}~\bibnamefont
  {Li}}, \bibinfo {author} {\bibfnamefont {J.-F.}\ \bibnamefont {Wyart}},
  \bibinfo {author} {\bibfnamefont {O.}~\bibnamefont {Dulieu}}, \bibinfo
  {author} {\bibfnamefont {S.}~\bibnamefont {Nascimb{\`e}ne}},\ and\ \bibinfo
  {author} {\bibfnamefont {M.}~\bibnamefont {Lepers}},\ }\bibfield  {title}
  {\bibinfo {title} {{Optical trapping of ultracold dysprosium atoms:
  transition probabilities, dynamic dipole polarizabilities and van der Waals
  C$_6$ coefficients}},\ }\href {https://doi.org/10.1088/1361-6455/50/1/014005}
  {\bibfield  {journal} {\bibinfo  {journal} {J. Phys. B: At. Mol. Opt. Phys.}\
  }\textbf {\bibinfo {volume} {50}},\ \bibinfo {pages} {014005} (\bibinfo
  {year} {2017})}\BibitemShut {NoStop}%
\bibitem [{\citenamefont {Safronova}\ \emph {et~al.}(2013)\citenamefont
  {Safronova}, \citenamefont {Safronova},\ and\ \citenamefont
  {Clark}}]{Safronova2013mwf}%
  \BibitemOpen
  \bibfield  {author} {\bibinfo {author} {\bibfnamefont {M.~S.}\ \bibnamefont
  {Safronova}}, \bibinfo {author} {\bibfnamefont {U.~I.}\ \bibnamefont
  {Safronova}},\ and\ \bibinfo {author} {\bibfnamefont {C.~W.}\ \bibnamefont
  {Clark}},\ }\bibfield  {title} {\bibinfo {title} {Magic wavelengths for
  optical cooling and trapping of potassium},\ }\href
  {https://doi.org/10.1103/PhysRevA.87.052504} {\bibfield  {journal} {\bibinfo
  {journal} {Phys. Rev. A}\ }\textbf {\bibinfo {volume} {87}},\ \bibinfo
  {pages} {052504} (\bibinfo {year} {2013})}\BibitemShut {NoStop}%
\bibitem [{\citenamefont {Kiruga}\ \emph {et~al.}(2025)\citenamefont {Kiruga},
  \citenamefont {Cheung}, \citenamefont {Filin}, \citenamefont {Barakhshan},
  \citenamefont {Bhosale}, \citenamefont {Badhan}, \citenamefont {Arora},
  \citenamefont {Eigenmann},\ and\ \citenamefont {Safronova}}]{Kiruga2025pfh}%
  \BibitemOpen
  \bibfield  {author} {\bibinfo {author} {\bibfnamefont {A.}~\bibnamefont
  {Kiruga}}, \bibinfo {author} {\bibfnamefont {C.}~\bibnamefont {Cheung}},
  \bibinfo {author} {\bibfnamefont {D.}~\bibnamefont {Filin}}, \bibinfo
  {author} {\bibfnamefont {P.}~\bibnamefont {Barakhshan}}, \bibinfo {author}
  {\bibfnamefont {A.}~\bibnamefont {Bhosale}}, \bibinfo {author} {\bibfnamefont
  {V.}~\bibnamefont {Badhan}}, \bibinfo {author} {\bibfnamefont
  {B.}~\bibnamefont {Arora}}, \bibinfo {author} {\bibfnamefont
  {R.}~\bibnamefont {Eigenmann}},\ and\ \bibinfo {author} {\bibfnamefont
  {M.~S.}\ \bibnamefont {Safronova}},\ }\href@noop {} {\bibinfo {title} {Portal
  for high-precision atomic data and computation}} (\bibinfo {year} {2025}),\
  \Eprint {https://arxiv.org/abs/2506.08170} {arXiv:2506.08170} \BibitemShut
  {NoStop}%
\bibitem [{\citenamefont {Partridge}\ \emph {et~al.}(2005)\citenamefont
  {Partridge}, \citenamefont {Strecker}, \citenamefont {Kamar}, \citenamefont
  {Jack},\ and\ \citenamefont {Hulet}}]{Partridge2005mpo}%
  \BibitemOpen
  \bibfield  {author} {\bibinfo {author} {\bibfnamefont {G.~B.}\ \bibnamefont
  {Partridge}}, \bibinfo {author} {\bibfnamefont {K.~E.}\ \bibnamefont
  {Strecker}}, \bibinfo {author} {\bibfnamefont {R.~I.}\ \bibnamefont {Kamar}},
  \bibinfo {author} {\bibfnamefont {M.~W.}\ \bibnamefont {Jack}},\ and\
  \bibinfo {author} {\bibfnamefont {R.~G.}\ \bibnamefont {Hulet}},\ }\bibfield
  {title} {\bibinfo {title} {{Molecular Probe of Pairing in the BEC-BCS
  Crossover}},\ }\href {https://doi.org/10.1103/PhysRevLett.95.020404}
  {\bibfield  {journal} {\bibinfo  {journal} {Phys. Rev. Lett.}\ }\textbf
  {\bibinfo {volume} {95}},\ \bibinfo {pages} {020404} (\bibinfo {year}
  {2005})}\BibitemShut {NoStop}%
\bibitem [{Note2()}]{Note2}%
  \BibitemOpen
  \bibinfo {note} {Because of the thermal spatial distribution in the trap, the
  molecules sample a mean intensity, which is typically 25$\protect \,\%$ below
  the peak intensity. Therefore our values of $\Gamma _{\protect \rm cc}$,
  which are simply calculated with the peak intensity, underestimate the true
  values. {In the special case of the 2002-nm trap light, interference effects
  may lead to locally increased trapping potentials. However, these effects
  remain small in comparison with the enormous variation of $\Gamma _{\protect
  \rm cc}$ observed in our experiments.}}\BibitemShut {Stop}%
\bibitem [{Note3()}]{Note3}%
  \BibitemOpen
  \bibinfo {note} {Examples for the threshold wavelength of bi-alkali molecules
  are \SI {1820}{\nano m} for KRb \cite {Ni2008ahp}, \SI {1890}{\nano m} for
  NaCs \cite {Warner2023ept}, \SI {2000}{\nano m} for RbCs \cite
  {Takekoshi2014uds}. For dimers involving other atoms: \SI {2200}{\nano m} for
  RbSr \cite {Zuchowski2014gae}, \SI {2660}{\nano m} for LiCr \cite
  {Finelli2024ula}.}\BibitemShut {Stop}%
\bibitem [{\citenamefont {Takekoshi}\ \emph {et~al.}(1998)\citenamefont
  {Takekoshi}, \citenamefont {Patterson},\ and\ \citenamefont
  {Knize}}]{Takekoshi1998ooo}%
  \BibitemOpen
  \bibfield  {author} {\bibinfo {author} {\bibfnamefont {T.}~\bibnamefont
  {Takekoshi}}, \bibinfo {author} {\bibfnamefont {B.~M.}\ \bibnamefont
  {Patterson}},\ and\ \bibinfo {author} {\bibfnamefont {R.~J.}\ \bibnamefont
  {Knize}},\ }\bibfield  {title} {\bibinfo {title} {Observation of optically
  trapped cold cesium molecules},\ }\href
  {https://doi.org/10.1103/PhysRevLett.81.5105} {\bibfield  {journal} {\bibinfo
   {journal} {Phys. Rev. Lett.}\ }\textbf {\bibinfo {volume} {81}},\ \bibinfo
  {pages} {5105} (\bibinfo {year} {1998})}\BibitemShut {NoStop}%
\bibitem [{\citenamefont {Cubizolles}\ \emph {et~al.}(2003)\citenamefont
  {Cubizolles}, \citenamefont {Bourdel}, \citenamefont {Kokkelmans},
  \citenamefont {Shlyapnikov},\ and\ \citenamefont
  {Salomon}}]{Cubizolles2003pol}%
  \BibitemOpen
  \bibfield  {author} {\bibinfo {author} {\bibfnamefont {J.}~\bibnamefont
  {Cubizolles}}, \bibinfo {author} {\bibfnamefont {T.}~\bibnamefont {Bourdel}},
  \bibinfo {author} {\bibfnamefont {S.~J. J. M.~F.}\ \bibnamefont
  {Kokkelmans}}, \bibinfo {author} {\bibfnamefont {G.~V.}\ \bibnamefont
  {Shlyapnikov}},\ and\ \bibinfo {author} {\bibfnamefont {C.}~\bibnamefont
  {Salomon}},\ }\bibfield  {title} {\bibinfo {title} {{Production of Long-Lived
  Ultracold Li$_2$ Molecules from a Fermi Gas}},\ }\href
  {https://doi.org/10.1103/PhysRevLett.91.240401} {\bibfield  {journal}
  {\bibinfo  {journal} {Phys. Rev. Lett.}\ }\textbf {\bibinfo {volume} {91}},\
  \bibinfo {pages} {240401} (\bibinfo {year} {2003})}\BibitemShut {NoStop}%
\bibitem [{\citenamefont {Jochim}\ \emph
  {et~al.}(2003{\natexlab{b}})\citenamefont {Jochim}, \citenamefont
  {Bartenstein}, \citenamefont {Altmeyer}, \citenamefont {Hendl}, \citenamefont
  {Chin}, \citenamefont {{Hecker Denschlag}},\ and\ \citenamefont
  {Grimm}}]{Jochim2003pgo}%
  \BibitemOpen
  \bibfield  {author} {\bibinfo {author} {\bibfnamefont {S.}~\bibnamefont
  {Jochim}}, \bibinfo {author} {\bibfnamefont {M.}~\bibnamefont {Bartenstein}},
  \bibinfo {author} {\bibfnamefont {A.}~\bibnamefont {Altmeyer}}, \bibinfo
  {author} {\bibfnamefont {G.}~\bibnamefont {Hendl}}, \bibinfo {author}
  {\bibfnamefont {C.}~\bibnamefont {Chin}}, \bibinfo {author} {\bibfnamefont
  {J.}~\bibnamefont {{Hecker Denschlag}}},\ and\ \bibinfo {author}
  {\bibfnamefont {R.}~\bibnamefont {Grimm}},\ }\bibfield  {title} {\bibinfo
  {title} {{Pure Gas of Optically Trapped Molecules Created from Fermionic
  Atoms}},\ }\href {https://doi.org/10.1103/PhysRevLett.91.240402} {\bibfield
  {journal} {\bibinfo  {journal} {Phys. Rev. Lett}\ }\textbf {\bibinfo {volume}
  {91}},\ \bibinfo {pages} {240402} (\bibinfo {year}
  {2003}{\natexlab{b}})}\BibitemShut {NoStop}%
\bibitem [{\citenamefont {Regal}\ \emph
  {et~al.}(2004{\natexlab{b}})\citenamefont {Regal}, \citenamefont {Greiner},\
  and\ \citenamefont {Jin}}]{Regal2004lom}%
  \BibitemOpen
  \bibfield  {author} {\bibinfo {author} {\bibfnamefont {C.~A.}\ \bibnamefont
  {Regal}}, \bibinfo {author} {\bibfnamefont {M.}~\bibnamefont {Greiner}},\
  and\ \bibinfo {author} {\bibfnamefont {D.~S.}\ \bibnamefont {Jin}},\
  }\bibfield  {title} {\bibinfo {title} {{Lifetime of Molecule-Atom Mixtures
  near a Feshbach Resonance in $^{40}$K}},\ }\href
  {https://doi.org/10.1103/PhysRevLett.92.083201} {\bibfield  {journal}
  {\bibinfo  {journal} {Phys. Rev. Lett.}\ }\textbf {\bibinfo {volume} {92}},\
  \bibinfo {eid} {083201} (\bibinfo {year} {2004}{\natexlab{b}})}\BibitemShut
  {NoStop}%
\bibitem [{\citenamefont {Gao}(2010)}]{Gao2010umf}%
  \BibitemOpen
  \bibfield  {author} {\bibinfo {author} {\bibfnamefont {B.}~\bibnamefont
  {Gao}},\ }\bibfield  {title} {\bibinfo {title} {Universal model for exoergic
  bimolecular reactions and inelastic processes},\ }\href
  {https://doi.org/10.1103/PhysRevLett.105.263203} {\bibfield  {journal}
  {\bibinfo  {journal} {Phys. Rev. Lett.}\ }\textbf {\bibinfo {volume} {105}},\
  \bibinfo {pages} {263203} (\bibinfo {year} {2010})}\BibitemShut {NoStop}%
\bibitem [{\citenamefont {Julienne}\ \emph {et~al.}(2011)\citenamefont
  {Julienne}, \citenamefont {Hanna},\ and\ \citenamefont
  {Idziaszek}}]{Julienne2011uuc}%
  \BibitemOpen
  \bibfield  {author} {\bibinfo {author} {\bibfnamefont {P.~S.}\ \bibnamefont
  {Julienne}}, \bibinfo {author} {\bibfnamefont {T.~M.}\ \bibnamefont
  {Hanna}},\ and\ \bibinfo {author} {\bibfnamefont {Z.}~\bibnamefont
  {Idziaszek}},\ }\bibfield  {title} {\bibinfo {title} {Universal ultracold
  collision rates for polar molecules of two alkali-metal atoms},\ }\href
  {https://doi.org/10.1039/C1CP21270B} {\bibfield  {journal} {\bibinfo
  {journal} {Phys. Chem. Chem. Phys.}\ }\textbf {\bibinfo {volume} {13}},\
  \bibinfo {pages} {19114} (\bibinfo {year} {2011})}\BibitemShut {NoStop}%
\bibitem [{\citenamefont {Qu\'em\'ener}\ \emph {et~al.}(2011)\citenamefont
  {Qu\'em\'ener}, \citenamefont {Bohn}, \citenamefont {Petrov},\ and\
  \citenamefont {Kotochigova}}]{Quenemer2011uiu}%
  \BibitemOpen
  \bibfield  {author} {\bibinfo {author} {\bibfnamefont {G.}~\bibnamefont
  {Qu\'em\'ener}}, \bibinfo {author} {\bibfnamefont {J.}~\bibnamefont {Bohn}},
  \bibinfo {author} {\bibfnamefont {A.}~\bibnamefont {Petrov}},\ and\ \bibinfo
  {author} {\bibfnamefont {S.}~\bibnamefont {Kotochigova}},\ }\bibfield
  {title} {\bibinfo {title} {Universalities in ultracold reactions of
  alkali-metal polar molecules},\ }\href
  {https://doi.org/10.1103/PhysRevA.84.062703} {\bibfield  {journal} {\bibinfo
  {journal} {Phys. Rev. A}\ }\textbf {\bibinfo {volume} {84}},\ \bibinfo
  {pages} {062703} (\bibinfo {year} {2011})}\BibitemShut {NoStop}%
\bibitem [{\citenamefont {Chin}\ and\ \citenamefont
  {Grimm}(2004)}]{Chin2004tea}%
  \BibitemOpen
  \bibfield  {author} {\bibinfo {author} {\bibfnamefont {C.}~\bibnamefont
  {Chin}}\ and\ \bibinfo {author} {\bibfnamefont {R.}~\bibnamefont {Grimm}},\
  }\bibfield  {title} {\bibinfo {title} {Thermal equilibrium and efficient
  evaporation of an ultracold atom-molecule mixture},\ }\href
  {https://doi.org/10.1103/PhysRevA.69.033612} {\bibfield  {journal} {\bibinfo
  {journal} {Phys. Rev. A}\ }\textbf {\bibinfo {volume} {69}},\ \bibinfo {eid}
  {033612} (\bibinfo {year} {2004})}\BibitemShut {NoStop}%
\bibitem [{\citenamefont {Marcelis}\ \emph {et~al.}(2008)\citenamefont
  {Marcelis}, \citenamefont {Kokkelmans}, \citenamefont {Shlyapnikov},\ and\
  \citenamefont {Petrov}}]{Marcelis2008cpo}%
  \BibitemOpen
  \bibfield  {author} {\bibinfo {author} {\bibfnamefont {B.}~\bibnamefont
  {Marcelis}}, \bibinfo {author} {\bibfnamefont {S.~J. J. M.~F.}\ \bibnamefont
  {Kokkelmans}}, \bibinfo {author} {\bibfnamefont {G.~V.}\ \bibnamefont
  {Shlyapnikov}},\ and\ \bibinfo {author} {\bibfnamefont {D.~S.}\ \bibnamefont
  {Petrov}},\ }\bibfield  {title} {\bibinfo {title} {Collisional properties of
  weakly bound heteronuclear dimers},\ }\href
  {https://doi.org/10.1103/PhysRevA.77.032707} {\bibfield  {journal} {\bibinfo
  {journal} {Phys. Rev. A}\ }\textbf {\bibinfo {volume} {77}},\ \bibinfo
  {pages} {032707} (\bibinfo {year} {2008})}\BibitemShut {NoStop}%
\bibitem [{\citenamefont {Ketterle}\ and\ \citenamefont {{van
  Druten}}(1996)}]{ketterle1996eco}%
  \BibitemOpen
  \bibfield  {author} {\bibinfo {author} {\bibfnamefont {W.}~\bibnamefont
  {Ketterle}}\ and\ \bibinfo {author} {\bibfnamefont {N.~J.}\ \bibnamefont
  {{van Druten}}},\ }\bibfield  {title} {\bibinfo {title} {{Evaporative Cooling
  of Trapped Atoms}},\ }\href {https://doi.org/10.1016/S1049-250X(08)60101-9}
  {\bibfield  {journal} {\bibinfo  {journal} {Adv. At. Mol. Opt. Phys.}\
  }\textbf {\bibinfo {volume} {37}},\ \bibinfo {pages} {181} (\bibinfo {year}
  {1996})}\BibitemShut {NoStop}%
\bibitem [{\citenamefont {Liu}\ \emph {et~al.}(2021)\citenamefont {Liu},
  \citenamefont {Yao}, \citenamefont {Chen}, \citenamefont {Wang},
  \citenamefont {Wang}, \citenamefont {Chen}, \citenamefont {Chen},
  \citenamefont {Levin},\ and\ \citenamefont {Pan}}]{Liu2021oot}%
  \BibitemOpen
  \bibfield  {author} {\bibinfo {author} {\bibfnamefont {X.-P.}\ \bibnamefont
  {Liu}}, \bibinfo {author} {\bibfnamefont {X.-C.}\ \bibnamefont {Yao}},
  \bibinfo {author} {\bibfnamefont {H.-Z.}\ \bibnamefont {Chen}}, \bibinfo
  {author} {\bibfnamefont {X.-Q.}\ \bibnamefont {Wang}}, \bibinfo {author}
  {\bibfnamefont {Y.-X.}\ \bibnamefont {Wang}}, \bibinfo {author}
  {\bibfnamefont {Y.-A.}\ \bibnamefont {Chen}}, \bibinfo {author}
  {\bibfnamefont {Q.}~\bibnamefont {Chen}}, \bibinfo {author} {\bibfnamefont
  {K.}~\bibnamefont {Levin}},\ and\ \bibinfo {author} {\bibfnamefont {J.-W.}\
  \bibnamefont {Pan}},\ }\bibfield  {title} {\bibinfo {title} {Observation of
  the density dependence of the closed-channel fraction of a {$^6$Li}
  superfluid},\ }\href {https://doi.org/10.1093/nsr/nwab226} {\bibfield
  {journal} {\bibinfo  {journal} {Natl. Sci. Rev.}\ }\textbf {\bibinfo {volume}
  {9}},\ \bibinfo {pages} {nwab226} (\bibinfo {year} {2021})}\BibitemShut
  {NoStop}%
\bibitem [{\citenamefont {J\"ager}\ and\ \citenamefont
  {Denschlag}(2024)}]{Jaeger2024ppm}%
  \BibitemOpen
  \bibfield  {author} {\bibinfo {author} {\bibfnamefont {M.}~\bibnamefont
  {J\"ager}}\ and\ \bibinfo {author} {\bibfnamefont {J.~H.}\ \bibnamefont
  {Denschlag}},\ }\bibfield  {title} {\bibinfo {title} {{Precise
  Photoexcitation Measurement of Tan's Contact in the Entire BCS-BEC
  Crossover}},\ }\href {https://doi.org/10.1103/PhysRevLett.132.263401}
  {\bibfield  {journal} {\bibinfo  {journal} {Phys. Rev. Lett.}\ }\textbf
  {\bibinfo {volume} {132}},\ \bibinfo {pages} {263401} (\bibinfo {year}
  {2024})}\BibitemShut {NoStop}%
\bibitem [{\citenamefont {Journeaux}\ \emph {et~al.}(2026)\citenamefont
  {Journeaux}, \citenamefont {Veschambre}, \citenamefont {Lecomte},
  \citenamefont {Uzan}, \citenamefont {Dalibard}, \citenamefont {Werner},
  \citenamefont {Petrov},\ and\ \citenamefont {Lopes}}]{Journeaux2026tbc}%
  \BibitemOpen
  \bibfield  {author} {\bibinfo {author} {\bibfnamefont {A.}~\bibnamefont
  {Journeaux}}, \bibinfo {author} {\bibfnamefont {J.}~\bibnamefont
  {Veschambre}}, \bibinfo {author} {\bibfnamefont {M.}~\bibnamefont {Lecomte}},
  \bibinfo {author} {\bibfnamefont {E.}~\bibnamefont {Uzan}}, \bibinfo {author}
  {\bibfnamefont {J.}~\bibnamefont {Dalibard}}, \bibinfo {author}
  {\bibfnamefont {F.}~\bibnamefont {Werner}}, \bibinfo {author} {\bibfnamefont
  {D.~S.}\ \bibnamefont {Petrov}},\ and\ \bibinfo {author} {\bibfnamefont
  {R.}~\bibnamefont {Lopes}},\ }\bibfield  {title} {\bibinfo {title} {{Two-Body
  Contact Dynamics in a Bose Gas near a Fano-Feshbach Resonance}},\ }\href
  {https://doi.org/10.1103/prf8-3q27} {\bibfield  {journal} {\bibinfo
  {journal} {Phys. Rev. Lett.}\ }\textbf {\bibinfo {volume} {136}},\ \bibinfo
  {pages} {083404} (\bibinfo {year} {2026})}\BibitemShut {NoStop}%
\bibitem {repository}%
  \BibitemOpen\href
  {https://doi.org/10.48323/bh0d4-qvq36} {10.48323/bh0d4-qvq36} \BibitemShut {NoStop}%
\bibitem [{\citenamefont {Warner}\ \emph {et~al.}(2023)\citenamefont {Warner},
  \citenamefont {Bigagli}, \citenamefont {Lam}, \citenamefont {Yuan},
  \citenamefont {Zhang}, \citenamefont {Stevenson},\ and\ \citenamefont
  {Will}}]{Warner2023ept}%
  \BibitemOpen
  \bibfield  {author} {\bibinfo {author} {\bibfnamefont {C.}~\bibnamefont
  {Warner}}, \bibinfo {author} {\bibfnamefont {N.}~\bibnamefont {Bigagli}},
  \bibinfo {author} {\bibfnamefont {A.~Z.}\ \bibnamefont {Lam}}, \bibinfo
  {author} {\bibfnamefont {W.}~\bibnamefont {Yuan}}, \bibinfo {author}
  {\bibfnamefont {S.}~\bibnamefont {Zhang}}, \bibinfo {author} {\bibfnamefont
  {I.}~\bibnamefont {Stevenson}},\ and\ \bibinfo {author} {\bibfnamefont
  {S.}~\bibnamefont {Will}},\ }\bibfield  {title} {\bibinfo {title} {{Efficient
  pathway to NaCs ground state molecules}},\ }\href
  {https://doi.org/10.1088/1367-2630/acd411} {\bibfield  {journal} {\bibinfo
  {journal} {New J. Phys.}\ }\textbf {\bibinfo {volume} {25}},\ \bibinfo
  {pages} {053036} (\bibinfo {year} {2023})}\BibitemShut {NoStop}%
\bibitem [{\citenamefont {Takekoshi}\ \emph {et~al.}(2014)\citenamefont
  {Takekoshi}, \citenamefont {Reichs\"ollner}, \citenamefont {Schindewolf},
  \citenamefont {Hutson}, \citenamefont {Le~Sueur}, \citenamefont {Dulieu},
  \citenamefont {Ferlaino}, \citenamefont {Grimm},\ and\ \citenamefont
  {N\"agerl}}]{Takekoshi2014uds}%
  \BibitemOpen
  \bibfield  {author} {\bibinfo {author} {\bibfnamefont {T.}~\bibnamefont
  {Takekoshi}}, \bibinfo {author} {\bibfnamefont {L.}~\bibnamefont
  {Reichs\"ollner}}, \bibinfo {author} {\bibfnamefont {A.}~\bibnamefont
  {Schindewolf}}, \bibinfo {author} {\bibfnamefont {J.~M.}\ \bibnamefont
  {Hutson}}, \bibinfo {author} {\bibfnamefont {C.~R.}\ \bibnamefont
  {Le~Sueur}}, \bibinfo {author} {\bibfnamefont {O.}~\bibnamefont {Dulieu}},
  \bibinfo {author} {\bibfnamefont {F.}~\bibnamefont {Ferlaino}}, \bibinfo
  {author} {\bibfnamefont {R.}~\bibnamefont {Grimm}},\ and\ \bibinfo {author}
  {\bibfnamefont {H.-C.}\ \bibnamefont {N\"agerl}},\ }\bibfield  {title}
  {\bibinfo {title} {{Ultracold Dense Samples of Dipolar RbCs Molecules in the
  Rovibrational and Hyperfine Ground State}},\ }\href
  {https://doi.org/10.1103/PhysRevLett.113.205301} {\bibfield  {journal}
  {\bibinfo  {journal} {Phys. Rev. Lett.}\ }\textbf {\bibinfo {volume} {113}},\
  \bibinfo {pages} {205301} (\bibinfo {year} {2014})}\BibitemShut {NoStop}%
\bibitem [{\citenamefont {\ifmmode~\dot{Z}\else \.{Z}\fi{}uchowski}\ \emph
  {et~al.}(2014)\citenamefont {\ifmmode~\dot{Z}\else \.{Z}\fi{}uchowski},
  \citenamefont {Gu\'erout},\ and\ \citenamefont {Dulieu}}]{Zuchowski2014gae}%
  \BibitemOpen
  \bibfield  {author} {\bibinfo {author} {\bibfnamefont {P.~S.}\ \bibnamefont
  {\ifmmode~\dot{Z}\else \.{Z}\fi{}uchowski}}, \bibinfo {author} {\bibfnamefont
  {R.}~\bibnamefont {Gu\'erout}},\ and\ \bibinfo {author} {\bibfnamefont
  {O.}~\bibnamefont {Dulieu}},\ }\bibfield  {title} {\bibinfo {title} {Ground-
  and excited-state properties of the polar and paramagnetic {RbSr} molecule: A
  comparative study},\ }\href {https://doi.org/10.1103/PhysRevA.90.012507}
  {\bibfield  {journal} {\bibinfo  {journal} {Phys. Rev. A}\ }\textbf {\bibinfo
  {volume} {90}},\ \bibinfo {pages} {012507} (\bibinfo {year}
  {2014})}\BibitemShut {NoStop}%
\end{thebibliography}
\end{document}